\documentclass[a4paper,fleqn,usenatbib]{mnras}

\usepackage{newtxtext,newtxmath}
\usepackage[T1]{fontenc}
\usepackage{ae,aecompl}

\usepackage{graphicx}	% Including figure files
\usepackage{amsmath}	% Advanced maths commands  %To put text mode inside math mode
\usepackage{subcaption}
\usepackage[flushleft]{threeparttable}
\usepackage[pdftex,usenames,dvipsnames]{xcolor}
\usepackage{natbib}
\usepackage{url}
\usepackage[none]{hyphenat}
\usepackage{times}
\usepackage{array}
\usepackage{lscape}
\usepackage{rotating}

\title[AGN outflows in dwarf galaxies]{The largest sample of AGN outflows in dwarf galaxies using DESI DR1}

\author[V. Rodríguez Morales et al.]{
V. Rodr\'{i}guez Morales,$^{1}$\thanks{E-mail: victorproyecto98@gmail.com}
M. Mezcua,$^{1,2}$
Ragadeepika Pucha,$^{3,4}$
C. Circosta,$^{5,6}$
S. Juneau,$^{7}$
M. Siudek,$^{1,8}$
\newauthor
S. Panda,$^{9}$
J. Aguilar,$^{10}$
S. Ahlen,$^{11}$
D. Bianchi,$^{12,13}$
D. Brooks,$^{5}$
T. Claybaugh,$^{10}$
A. Cuceu,$^{10}$
A. de la Macorra,$^{14}$
\newauthor
B. Dey,$^{15,16}$
P. Doel,$^{5}$
J.E. Forero-Romero,$^{17,18}$
S. Gontcho A Gontcho,$^{19}$
G. Gutierrez,$^{20}$
J. Guy,$^{10}$
C. Hahn,$^{21}$
\newauthor
R. Joyce,$^{7}$
A. Kremin,$^{10}$
A. Lambert,$^{10}$
M. Landriau,$^{10}$
L. Le Guillou,$^{22}$
M. Manera,$^{23,24}$
P. Martini,$^{25,26,27}$
\newauthor
A. Meisner,$^{7}$
R. Miquel,$^{24,28}$
J. Moustakas,$^{29}$
W.J. Percival,$^{30,31,32}$
C. Poppett,$^{7,33,34}$
F. Prada,$^{35}$
I. P\'{e}rez-R\`{a}fols,$^{36}$
\newauthor
G. Rossi,$^{37}$
E. Sanchez,$^{38}$
D. Schlegel,$^{10}$
M. Schubnell,$^{39,40}$
D. Sprayberry,$^{7}$
G. Tarl\'{e},$^{40}$
B.A. Weaver,$^{7}$
\newauthor
and H. Zoo$^{41}$
}

\date{Accepted XXX. Received YYY; in original form ZZZ}
\pubyear{XXXX}

\begin{document}
\label{firstpage}
\pagerange{\pageref{firstpage}--\pageref{lastpage}}
\maketitle

\begin{abstract}
   In the last decade, the presence of active galactic nuclei (AGN) outflows and feedback in dwarf galaxies ($\mathrm{M_\ast}$\,$<$\,$10^{10}\mathrm{M}_\odot$) has gained ground over supernova (SN) feedback as the main mechanism regulating star formation. In this work, we perform the first systematic search for AGN outflows in dwarf galaxies using the Dark Energy Spectroscopic Instrument Data Release 1 (DESI DR1). From $\sim$ 7 million galaxies at z$<$0.45, we identify ionized outflows through the detection of broad components in the [OIII]$\lambda5007$\AA \  emission line. Galaxies are divided into dwarf and massive systems. Then, using emission-line diagnostic diagrams, we classify as star forming or AGN. We identify 1,502 AGN dwarf galaxies with outflow signatures. Comparing the distributions of star forming and AGN galaxies with outflows, we find that, among the 1,502 AGN dwarf galaxies with outflow signatures, AGN are the most likely drivers of the observed outflows in $\sim$83$\%$ of those with W$_{80}$ velocity $>250$ km s$^{-1}$. This constitutes the largest statistical sample of AGN outflows in dwarf galaxies to date. In massive galaxies, AGN dominance occurs above W$_{80}>350$ km s$^{-1}$. Therefore, two new velocity thresholds are proposed for identifying AGN-driven outflows in dwarf and massive galaxies. Besides, we find that outflows in dwarf galaxies are more likely to escape the dark matter halo than those in massive galaxies, allowing gas to be redistributed from the inner to the outer regions. This suggests that AGN outflows may have a major impact on dwarf galaxies.
\end{abstract}


\section{Introduction}

Supermassive black holes (SMBHs; black hole mass $\mathrm{M_{BH}}$\,$>$\,$10^6\mathrm{M_\odot}$) are located at the center of nearly all massive galaxies (stellar mass $\mathrm{M_\ast}$\,$>$\,$10^{10}\mathrm{M_\odot}$). SMBHs are primarily formed via mergers and accretion from smaller building blocks known as `seed' black holes (BHs). The growth of SMBHs is connected with the evolution of their host galaxy through the accretion of matter that powers the active galactic nuclei (AGN). This is known as BH-galaxy co-evolution (\citealt{kormendy2013coevolution}; \citealt{zhuang2023evolutionary}; \citealt{capelo2024black}). Although the technological capabilities of directly detecting seed BHs are not yet developed enough, dwarf galaxies ($\mathrm{M_\ast}$\,$<$\,$10^{10}\mathrm{M}_\odot$) in the local Universe are thought to host intermediate-mass BHs (IMBHs; $100\mathrm{M_\odot}$\,$<$\,$\mathrm{M_{BH}}$\,$<$\,$10^6\mathrm{M_\odot}$) that resemble those seeds formed in the early Universe (z $\sim$ 15; e.g. \citealt{mezcua2017observational}; \citealt{greene2020intermediate}; \citealt{mezcua2023overmassive}), as some may not have experienced significant evolution. Consequently, understanding how IMBHs in dwarf galaxies interact with their host galaxies through feedback is essential. Dwarf galaxies may offer a unique window into the physical processes that regulated the earliest stages of BH growth, and provide key insights into how seed BHs were fuelled before becoming the SMBHs we observe today.

In recent years, thousands of dwarf galaxies hosting AGN have been revealed (\citealt{reines2013dwarf}; \citealt{moran2014black}; \citealt{polimera2022resolve}; \citealt{siudek2023environment}; \citealt{mezcua2024manga}; \citealt{pucha2025tripling}) thanks to spectroscopic surveys such as the Sloan Digital Sky Survey (SDSS; \citealt{york2000sloan}), the VIMOS Public Extragalactic Redshift Survey (VIPERS; \citealt{scodeggio2018vimos}), the Galaxy and Mass Assembly (GAMA; \citealt{driver2009gama}) survey, and the Dark Energy Spectroscopic Instrument (DESI; \citealt{aghamousa2016desi, abareshi2022overview}). This has been mostly based on emission-line diagnostic diagrams such as the Baldwin, Phillips, and Terlevich (BPT; \citealt{baldwin1981classification}) diagram. Furthermore, tens of AGN in dwarf galaxies have been found in other wavelength regimes using different methods like the infrared color-color diagrams (\citealt{sartori2015search}; \citealt{kaviraj2019agn}; \citealt{lupi2020difficulties}), radio observations (\citealt{mezcua2019radio}; \citealt{reines2020new}; \citealt{davis2022radio}; \citealt{eberhard2025dwarf}; \citealt{flores2025manga}), variability (\citealt{baldassare2018identifying, baldassare2020search}; \citealt{burke2022dwarf}; \citealt{wasleske2022variable}; \citealt{ward2022variability}) or X-ray emission (\citealt{lemons2015x}; \citealt{mezcua2016population, mezcua2018intermediate}; \citealt{birchall2020x}; \citealt{sacchi2024x}).

With the new detections of AGN in dwarf galaxies, a new paradigm about how these galaxies evolved, and which feedback mechanisms are involved, has been opened. In the common theoretical models, reionization and SN feedback are considered the main mechanisms for regulating the SF (e.g. \citealt{dekel1986origin}; \citealt{efstathiou1992suppressing}; \citealt{mori2002early}; \citealt{okamoto2008mass}), and thus shaping the evolutionary pathways of dwarf galaxies. However, it is not clear if these processes are sufficient to regulate star formation (SF). Some works point out the need for additional physical processes (e.g.  photoionization, radiation pressure, or stellar winds) for supernovae (SNe) to effectively drive outflows and couple to the interstellar medium (ISM; \citealt{hopkins2014galaxies}; \citealt{kimm2015towards}; \citealt{smith2019cosmological}). In addition, recent theoretical models have identified physical conditions under which AGN in dwarf galaxies can efficiently accrete. With these models, AGN are found to have an impact on the host galaxy by driving outflows and suppressing star formation (e.g. \citealt{barai2019intermediate}; \citealt{sharma2020black, sharma2022hidden, sharma2023active}; \citealt{koudmani2019fast, koudmani2021little, koudmani2022two}; \citealt{wellons2023exploring}; \citealt{arjona2024role}), although other models suggest that AGN feedback enhances the star formation (e.g. \citealt{su2025positive}). Recent simulations also study the possible impact of AGN feedback on dark matter distributions in order to explain the diversity of rotation curves observed in dwarf galaxies (\citealt{koudmani2025diverse}; \citealt{jackson2025diversity}). In the end, the improving of baryonic physics modelling may largely alleviate the different controversies known as the `dwarf galaxies problems' (e.g. the missing satellite problem, the too-big-to-fail problem, the cusp-core problem; see \citealt{sales2022baryonic} for a recent review).

%\raga{Observational studies have furthermore reported evidence of AGN feedback and outflows in the low-mass regime. \cite{penny2018sdss} reported the first evidence of AGN feedback in a subset of 69 quenched dwarf galaxies. \textbf{Also, the detection of AGN-driven outflows in dwarf galaxies suggests an impact on these galaxies carried by the AGN: \cite{manzano2019agn} reported the detection of six dwarf galaxies with AGN outflow signatures based on the detection of a broad component in the [OIII] doublet. \cite{liu2020integral} studied eight of the AGN dwarf galaxies from the \cite{manzano2019agn} sample, analyzing them with integral field spectroscopic data, detecting outflowing gas in seven of them. \cite{zheng2023escaping} detected an AGN outflow able to escape from the galaxy in a dwarf galaxy hosting an IMBH. \cite{wang2024rubies} reported a dwarf galaxy with AGN signatures at redshift z $=3.1$ with outflow signatures based on an absorption feature in the wings of [HeI]$\lambda10839$ \AA \ .\cite{salehirad2025ionized} reported 11 AGN dwarf galaxies with fast outflows driven by an AGN. \cite{morales2025manga} reported 13 AGN dwarf galaxies with AGN outflows and studied their kinematic and energetic properties. \cite{ivey2025exploring} detected two AGN dwarf galaxies with outflow signatures at z $\sim 7.6$}. However, the current sample of dwarf galaxies with confirmed AGN outflows remains too small to enable a statistical understanding of how these outflows regulate SF and thus affect dwarf galaxy evolution.}

Furthermore, observational evidence for AGN feedback and AGN outflows in dwarf galaxies has grown in recent years. Several studies have reported the presence of ionized outflows based on broad emission-line components, together with spatially resolved kinematics in some cases (e.g. \citealt{penny2018sdss}; \citealt{manzano2019agn}; \citealt{liu2020integral}; \citealt{zheng2023escaping}; \citealt{wang2025rubies}; \citealt{salehirad2025ionized}; \citealt{rodriguez2025manga}; \citealt{ivey2026exploring}). Some of these detections indicate that the outflowing gas may escape the shallow gravitational potential of these systems, potentially affecting the SF of their host galaxies. However, the detections are typically based on small samples or individual objects and span a wide range of redshifts. As a result, the current census of dwarf galaxies with confirmed AGN outflows remains limited, preventing robust statistical constraints on the role of AGN feedback in regulating SF and shaping dwarf galaxy evolution.

In this paper, we use optical spectroscopic data from the DESI Data Release 1 (DR1; \citealt{desi2026data}) to identify AGN outflows in dwarf galaxies and present the largest sample to date. We use the Python-based emission-line fitting code \texttt{EmFit}\footnote{\url{https://github.com/Ragadeepika-Pucha/EmFit}} (\citealt{pucha2025tripling,pucha2026new}) to identify additional components in the [OIII]$\lambda$5007 \AA \ emission line in galaxies. We compute the kinematic and energetic properties of these outflows and compare them to those of massive galaxies. We also compare the outflow velocity in AGN dwarf galaxies with that in non-AGN dwarf galaxies that present outflow signatures. By estimating the escape velocity, the coupling efficiency, and the mass-loading factor we investigate whether these outflows could affect the SF in the host galaxy. In Sect. \ref{sect:2}, we describe the sample selection and the estimation of its kinematic and energetic properties. In Sect. \ref{sect:3}, we present the main results of this work and discuss their implications. Final conclusions are drawn in Sect. \ref{sect:4}. Throughout the paper, a $\Lambda$CDM cosmology is adopted with $\mathrm{H_0 = 73 \ km \ s^{-1}}$, $\Omega_{\mathrm{M}}=0.27$.

%IN SECTION 2 we study an initial sample of 7427711 galaxies (both massive and dwarves, AGN and non-AGN)  at redshift z $<0.45$ taken from DESI DR1 and explain the different criteria and selection cuts applied to reduce our sample into a confident sample of AGN dwarf galaxies with outflows signatures, in addition to the equivalent control sample of massive galaxies. In this section, we also calculate the kinematic and energetic properties of the outflows. 

\begin{figure}
\centering
\includegraphics[width=0.5\textwidth]{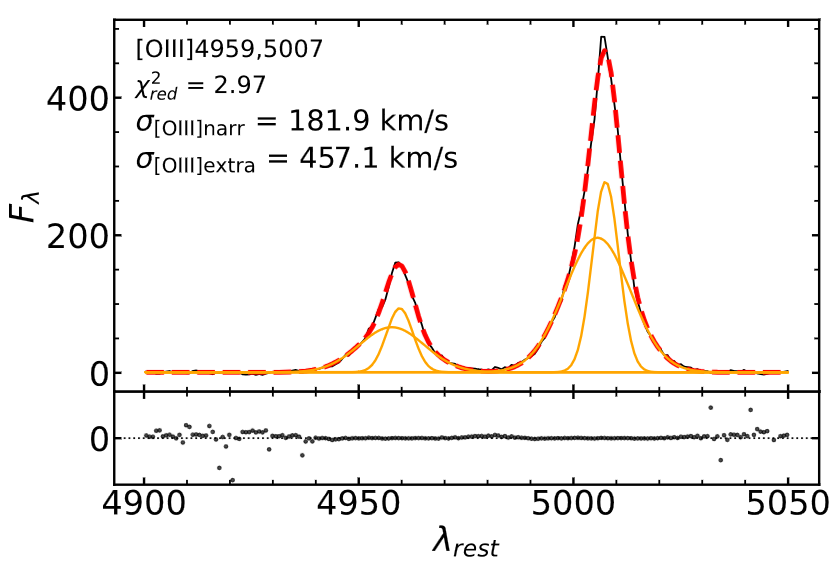} 
\caption{[OIII]$\lambda\lambda$ 4959,5007 \AA \  emission line profile of the galaxy with TARGETID 2851126508519424 fitted with \texttt{EmFit} (\citealt{pucha2025tripling}). The spectrum is shown in black, the red line shows the best-fit model to the continuum-subtracted emission line spectrum. The narrow and the broad component are plotted in orange. The reduced $\chi^2$ of the fit is shown in the upper-left corner. Residuals are shown in the bottom.}
\label{figure:broad_comp}
\end{figure}

\begin{figure*}
\centering
\includegraphics[width=\textwidth]{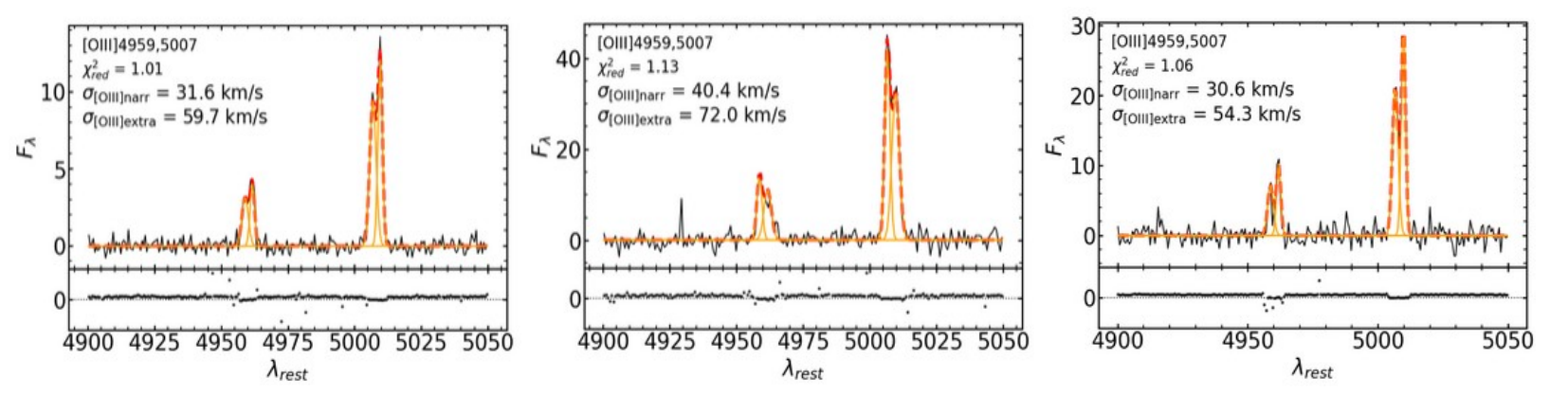} 
\caption{Same caption as Fig. \ref{figure:broad_comp} for the galaxies with TARGETID 39627938072629929, 39627473901586204, and 39627798951757773 (right, middle and left, respectively) classified as double-peaked based on the $|\mathrm{v_{off}}|-\sigma_\mathrm{[OIII]extra}>100$ km s$^{-1}$ condition.} 
\label{figure:double-peaked}
\end{figure*}

\section{Sample selection and methods}
\label{sect:2}

DESI is a 5000-fiber multi-object spectroscopic survey operating on the Mayall 4-meter telescope at Kitt Peak National Observatory (\citealt{abareshi2022overview}). DESI covers a spectral range of 3600-9800 \AA \  with a resolution of R $=$ 2000-5500, with the ability to obtain simultaneous spectra of almost 5000 objects over a $\sim$ 3 deg$^2$ field (\citealt{aghamousa2016desi}; \citealt{miller2024optical}; \citealt{poppett2024overview}). DESI will cover an area of $\approx$ 17,000 deg$^2$ in eight years, obtaining approximately 63 million spectra of galaxies and quasars (\citealt{hahn2023desi}; \citealt{raichoor2023target}; \citealt{chaussidon2023target}; \citealt{schlafly2023survey}). DESI DR1 was publicly released in March 2025, including spectra for $\sim$ 14.6 million galaxies and quasars (\citealt{adame2025desi}; \citealt{desi2026data}).

The spectra obtained from DESI were reduced using the DESI spectroscopic pipeline (\citealt{guy2023spectroscopic}), and the redshift for each source is provided by the \texttt{Redrock\footnote{https://github.com/desihub/redrock}} pipeline (\citealt{brodzeller2023performance}; \citealt{anand2024archetype}; Bailey et al. in prep.) in addition to its uncertainty (\texttt{ZERR}), a redshift warning bitmask (\texttt{ZWARN}) and a spectral type (\texttt{SPECTYPE}). From DESI DR1, we selected 7,378,347 low-redshift ($\mathrm{z}$\,$<$\,$0.45$) objects (\texttt{SPECTYPE}$==$\texttt{GALAXY} or \texttt{QSO}) with good spectra\footnote{Redshift warning \texttt{ZWARN}=0 or 4}, photometric information and emission-line measurements (see Sect. \ref{sect:emiision-line fitting}). These cuts are fully described in \cite{pucha2026new}.

\subsection{Stellar mass cut}
\label{sect:mass}

We considered stellar masses from the DESI DR1 Value Added Catalogue (VAC) of physical properties\footnote{\url{https://data.desi.lbl.gov/doc/releases/dr1/vac/cigale/}} (\citealt{siudek2024value}; Siudek et al. in prep.). The stellar masses in this catalogue were computed by modelling spectral energy distributions (SED) using Code Investigating GALaxy Emission (CIGALE; \citealt{boquien2019cigale}). The SED fitting used \cite{bruzual2003stellar} single stellar population (SSP) models adopting a \cite{chabrier2003galactic} initial mass function (IMF) to build a stellar component. It included nebular emission models pre-computed with \texttt{CLOUDY} (\citealt{ferland2017rev}) and dust attenuation based on the \cite{calzetti2000dust} attenuation curve. It also accounted for AGN contribution, which was modelled using \cite{fritz2006revisiting} model that covers from the ultraviolet to infrared and assumes that the central engine is surrounded by smoothly distributed dust in the AGN torus. Using the spectroscopic redshift of each galaxy, CIGALE performed simultaneous fit to the AGN and the galaxy component across all available photometric bands: ground-based optical and near-infrared photometry (i.e. $g, r, z$) from the Legacy Imaging Survey (\citealt{dey2019overview}), supplemented with observations from mid-infrared bands at 3.4, 4.6, 12 and 22 $\mu$m provided by the Wide-field Infrared Survey Explorer (WISE; \citealt{wright2010wide}). To select secure fits, we excluded those with $\chi^2_r$\,$>$\,$17$ and stellar mass $\mathrm{M}_\ast$\,$<$\,$10^5\mathrm{M}_\odot$ (corresponding with a $\sim 5\%$ of the sample), reducing the initial sample to 7,034,924 galaxies. For a more detailed description of the SED fitting see \cite{siudek2024value}.  

We applied a stellar mass cut to divide the sample into dwarf and massive galaxies. Following earlier studies, we considered galaxies with $\mathrm{M_\ast}$\,$\leq$\,$10^{10}\mathrm{M_\odot}$ as dwarf galaxies throughout this paper (e.g. \citealt{manzano2020active}; \citealt{siudek2023environment}; \citealt{bichang2024}; \citealt{erostegui2025}; \citealt{rodriguez2025manga}), whereas galaxies with $\mathrm{M_\ast}$\,$>$\,$10^{10} \mathrm{M_\odot}$ were considered massive. From the sample of 7,034,924 galaxies, $34\%$ are dwarf galaxies ($\sim 2.4$ million), while $66\%$ are massive galaxies ($\sim 4.6$ million).

\begin{figure*}
\centering
\includegraphics[width=\textwidth]{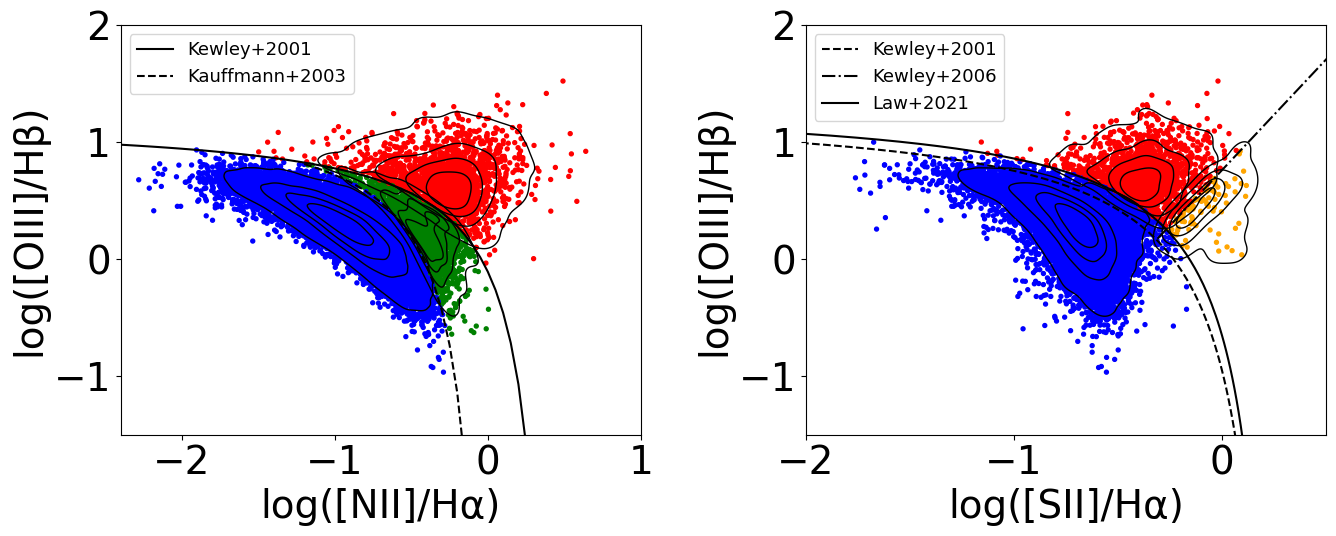} 
\caption{[NII]-BPT (left) and [SII]-BPT (right) diagnostic diagram of 22424 dwarf galaxies with observed outflow signatures used to distinguish the ionization source of each galaxy: AGN (red), SF (blue), Composite (green) and LINER (orange). For the [NII]-BPT \protect\cite{kewley2001theoretical} and \protect\cite{kauffmann2003host} demarcation lines are shown (solid and dashed, respectively). For the [SII]-BPT \protect\cite{kewley2001theoretical}, \protect\cite{kewley2006host} and \protect\cite{law2021sdss} demarcation lines are shown (dashed, dash-dotted, and solid, respectively).} 
\label{figure:BPT}
\end{figure*}

\subsection{Emission-line fitting and [OIII]$\lambda$5007 emission line cut}
\label{sect:emiision-line fitting}

We relied on the VAC of emission-line measurements at $\mathrm{z}$\,$<$\,$0.45$ obtained with the emission-line fitting code \texttt{EmFit} (EmFit v2.3 \footnote{\url{https://data.desi.lbl.gov/public/dr1/vac/dr1/emfit}}; \citealt{pucha2025tripling, pucha2026new}). The code used a non-linear least squares fitting algorithm to fit eight emission lines:  H$\beta \ \lambda4861$~\AA, [OIII]$\lambda\lambda$4959, 5007~\AA, [NII]$\lambda\lambda$6548, 6584~\AA, H$\alpha \ \lambda$6563~\AA ,  and [SII]$\lambda\lambda$6717, 6731~\AA. This fitting algorithm tested for the presence of outflow component independently for the H$\beta$, [OIII], H$\alpha$, and [SII] emission lines. \texttt{EmFit} further tests for the presence of broad Balmer components. The [SII] profile was used as a template for fitting the [NII], H$\alpha$, and H$\beta$ emission lines. The [OIII] doublet was fitted separately, fixing the relative redshifts of the peaks, the amplitude ratio of [OIII]$\lambda$5007/[OIII]$\lambda$4959 to 2.98, and adding the condition that the widths of the two lines were equal in the velocity space. The number of components of H$\beta$ was fixed to be the same as H$\alpha$, being the redshifts and the widths of these components fixed to the respective H$\alpha$ components in the velocity space. For the sources flagged as extreme broad-line (EBL; i.e. sources where the broad component of H$\alpha$ extends up to the [SII] region), the redshifts and widths of the of the H$\beta$ components was fixed to scale with H$\alpha$ components. Further details related to the fitting methodology are described in Section 3.2 of \cite{pucha2025tripling}.

To identify galaxies wihth outflows, we selected sources with a broad component in the [OIII] emission line (see Fig. \ref{figure:broad_comp}). To remove non-physical fits we required: (i) an amplitude-over-noise (AoN; defined as the amplitude over the noise of the continuum) for the narrow component in the [OIII] emission line greater than 10 (AoN>10); (ii) a signal-to-noise ratio (SNR), defined as the flux of the outflow component divided by its error, greater than five (SNR $>5$), and (iii) the velocity dispersion of the broad component to be less than 2000 km s$^{-1}$ ($\sigma_b <2000$ km s$^{-1}$). Targets with higher $\sigma_b$ typically show the broad component of the [OIII]$\lambda$5007 blended with its counterpart in the [OIII]$\lambda4959$ emission line, contaminating the fitting. In addition, in some cases such a large component may be the result of fitting noise. Once these quality cuts were applied, we find 57,563 galaxies with outflow signatures. We did a visual inspection by randomly selecting a thousand galaxies out of the 57,563  to ensure these cuts provide a confident sample of galaxies with outflow signatures. Out of the 57,563 galaxies, 25,822 are dwarf galaxies ($45\%$), and 31,741 are massive galaxies ($55\%$).

\subsection{Double-peaked emission}
\label{sect: double-peaked}

An additional component in the [OIII] emission line does not always imply an outflow component, as it may also reflect a secondary narrow feature in a double-peaked emission-line galaxy (see Fig. \ref{figure:double-peaked}). The origin of the double-peaked emission has various causes: a compact rotating disc, two nuclei, or a merger scenario (e.g. \citealt{comerford2011chandra, comerford2015merger}; \citealt{maschmann2023origin}). In addition, AGN outflows can also produce this double-peaked emission (\citealt{muller2015origin}; \citealt{nevin2016origin}; \citealt{comerford2018origin}). Identifying the possible origin of the double-peaked emission was outside the scope of this work. Therefore, we excluded all the galaxies that exhibite double-peaked emission to avoid contaminating the sample with other potential origins. 

To identify the double-peaked emission, we looked if the extra component in the [OIII]$\lambda 5007$ \AA \ line had an amplitude and width similar to those of the narrow component. If the extra component was an outflow component, we would expect the opposite behaviour: a lower amplitude and a greater width. Thus, the ratio of the amplitude and the width between the extra and the narrow components were defined as:

\begin{equation}
\begin{aligned}
    \mathrm{AR_{[OIII]}} &=\frac{\mathrm{Amplitude_{[OIII]extra}}}{\mathrm{Amplitude_{[OIII]narr}}}\\  
    \mathrm{SR_{[OIII]}} &=\frac{\sigma_{\mathrm{[OIII]extra}}}{\sigma_{\mathrm{[OIII]narr}}} 
\end{aligned}
\end{equation}

where $\sigma_{\mathrm{[OIII]}}$ is the velocity dispersion, and `extra' and `narr' refer to the extra and the narrow component, respectively. \cite{pucha2025tripling} found an optimum cut for identifying the double-peaked emission in the [OIII] emission line based on visual inspection: $\mathrm{log(AR_{[OIII]})>-0.09}$ or $\mathrm{log(SR_{[OIII]}) < 0.2}$. In this work, we included an extra condition to identify these sources: $|\mathrm{v_{off}}|-\sigma_{\mathrm{[OIII]extra}}>100$ km s$^{-1}$, where $\mathrm{v_{off}}$ is the velocity offset between the extra and the narrow component. This condition was also based on visual inspection, inspired by the fact that the shift of an extra narrow component should be greater than its width. We classified a source as double-peaked if at least one of the three conditions was met. In Fig. \ref{figure:double-peaked}, we showed examples of three double-peaked emission-line sources selected only by the extra condition derived in this work. We found 3,459 (13$\%$) and 7,023 (22$\%$) dwarf and massive galaxies, respectively, with double-peaked emission that were excluded from the sample, leaving a set of 22,363 dwarf galaxies and 24,718 massive galaxies with outflow signatures.

\subsection{AGN selection}
\label{sect:agn}
To determine whether the outflows coincide with the presence of an AGN, we searched for AGN ionization signatures based on two optical emission-line diagnostics: [OIII]$\lambda5007$/H$\beta$ versus [NII]$\lambda6583$/H$\alpha$ (henceforth [NII]-BPT) and [OIII]$\lambda5007$/H$\beta$ versus [SII]$\lambda\lambda6717,6731$/H$\alpha$ (henceforth [SII]-BPT; \citealt{baldwin1981classification}; \citealt{veilleux1987spectral}). These diagrams only consider the narrow component of the emission lines. The [NII]-BPT and [SII]-BPT are biased toward solar and supersolar metallicities, and therefore miss a fraction of low-metallicity galaxies hosting AGN. This limitation can be circumvented by using the [OI]-BPT, which is insensitive to metallicity (e.g. \citealt{polimera2022resolve}; \citealt{mezcua2024manga}). However, in this paper we did not use the [OI]-BPT, as \texttt{EmFit} does not yet fit the [OI]$\lambda$6300 emission line.

For the [NII]-BPT we used the \cite{kewley2001theoretical} and \cite{kauffmann2003host} demarcation lines, which differentiate between AGN, SF and Composite (i.e. mixture of both AGN and SF, see left panel of Fig. \ref{figure:BPT}) ionization contributions. For the [SII]-BPT, we used the  \cite{law2021sdss} demarcation line instead of the \cite{kewley2001theoretical} line to separate between the SF and Seyfert ionization, as the one from \cite{kewley2001theoretical} may overestimate the AGN fraction in dwarf galaxies (\citealt{pucha2025tripling}). In addition, for the [SII]-BPT, we used the \cite{kewley2006host} line to distinguish between sources with Seyfert ionization and LINER\footnote{Low-ionization emission line region} ionization (see right panel of Fig. \ref{figure:BPT}). We required that the emission lines used in the BPT diagrams have a SNR greater than five (SNR$_\mathrm{EL}>5$). Furthermore, to avoid an over-detection of AGN galaxies, we accounted for the [NII]$\lambda$6583, [SII]$\lambda\lambda$6717,6731, H$\alpha$, and H$\beta$ double-peaked emission, based on \cite{pucha2025tripling} criteria. Therefore, we considered the flux  of these emission line as the sum of the two narrow components contributions when the double-peaked emission line was detected.\footnote{Although the BPT diagrams do not use the [OIII] emission line, we remind that for this emission line the double-peaked sources were already removed from the sample (see Sect. \ref{sect: double-peaked}).} 

We adopted \cite{mezcua2024manga} scheme to classify the source of ionization of each dwarf galaxy as: 

\begin{itemize}
\item{\textit{AGN} if AGN or Composite in the [NII]-BPT and Seyfert in the [SII]-BPT.}\\
\item{\textit{AGN-LINER} if AGN in the [NII]-BPT and SF or LINER in the [SII]-BPT.}\\
\item{\textit{Composite} if Composite in the [NII]-BPT and SF or LINER in the [SII]-BPT.}\\
\item{\textit{SF-AGN} if SF in the [NII]-BPT and AGN in the [SII]-BPT.}\\
\item{\textit{LINER}  if SF in the [NII]-BPT and LINER in the [SII]-BPT.}\\
\item{\textit{SF} if SF in both BPTs.}\\
\end{itemize}  

We selected those dwarf and massive galaxies classified as AGN, reducing the sample from 22,363 to 1,502 and from 24,718 to 10,855 AGN dwarf and massive galaxies with outflow signatures, respectively.

In order to select a control sample of non-AGN dwarf and massive galaxies with outflow signatures, we selected those galaxies classified as SF based on the [NII] and [SII]-BPT. We further required that they did not show any AGN signatures in other diagnostics (e.g. the H$\alpha$ equivalent width versus [NII]/H$\alpha$ line ratio diagram from \citealt{fernandes2010alternative}, [NeV]$\lambda 3426$ \AA \ detection, or WISE diagnostics from \citealt{jarrett2011spitzer}; \citealt{mateos2012using}; \citealt{stern2012mid}; \citealt{assef2018wise}; \citealt{yao2020galaxy}; \citealt{hviding2022new}), based on the AGN/Galaxy Classification Value-Added Catalogue for DESI DR1\footnote{\url{https://data.desi.lbl.gov/doc/releases/dr1/vac/agngal/}} (Juneau et al. in prep.), to avoid contamination. Also, for the control sample of non-AGN dwarf galaxies with outflow signatures, we matched in stellar mass by selecting objects lying between the 10th and 90th percentiles of the stellar-mass distribution of  AGN dwarf galaxies with outflow signatures. This resulted in a control sample of 3,651 non-AGN dwarf galaxies with comparable global properties to the AGN sample, with median values of $\mathrm{logM_\ast=}$ 9.6 and 9.8, $\mathrm{logL_{OIII}}=$ 41.6 and 41.7, and redshift  z $\sim$ 0.2, for the non-AGN and AGN sample, respectively. For massive galaxies, we divided the AGN sample with outflow signatures into two stellar-mass bins ($10^{10}\mathrm{M}_\odot$ $\leq$ M$_\ast \leq$ $10^{10.5}\mathrm{M}_\odot$ and M$_\ast>$ $10^{10.5}\mathrm{M}_\odot$, consisting of 3,877 and 6,978 galaxies, respectively ). The first bin was then compared to a control sample of 346 non-AGN massive galaxies with outflow signatures spanning the same stellar-mass range ($10^{10}\mathrm{M}_\odot$ $\leq$ M$_\ast \leq$ $10^{10.5}\mathrm{M}_\odot$). These two massive galaxy samples also exhibited comparable global properties in redshift and [OIII] luminosity, ensuring a consistent comparison between them.

\subsection{Kinematic and energetic properties of the outflows}
\label{sect:properties}

\begin{figure*}
    \centering
    \includegraphics[width=\textwidth]{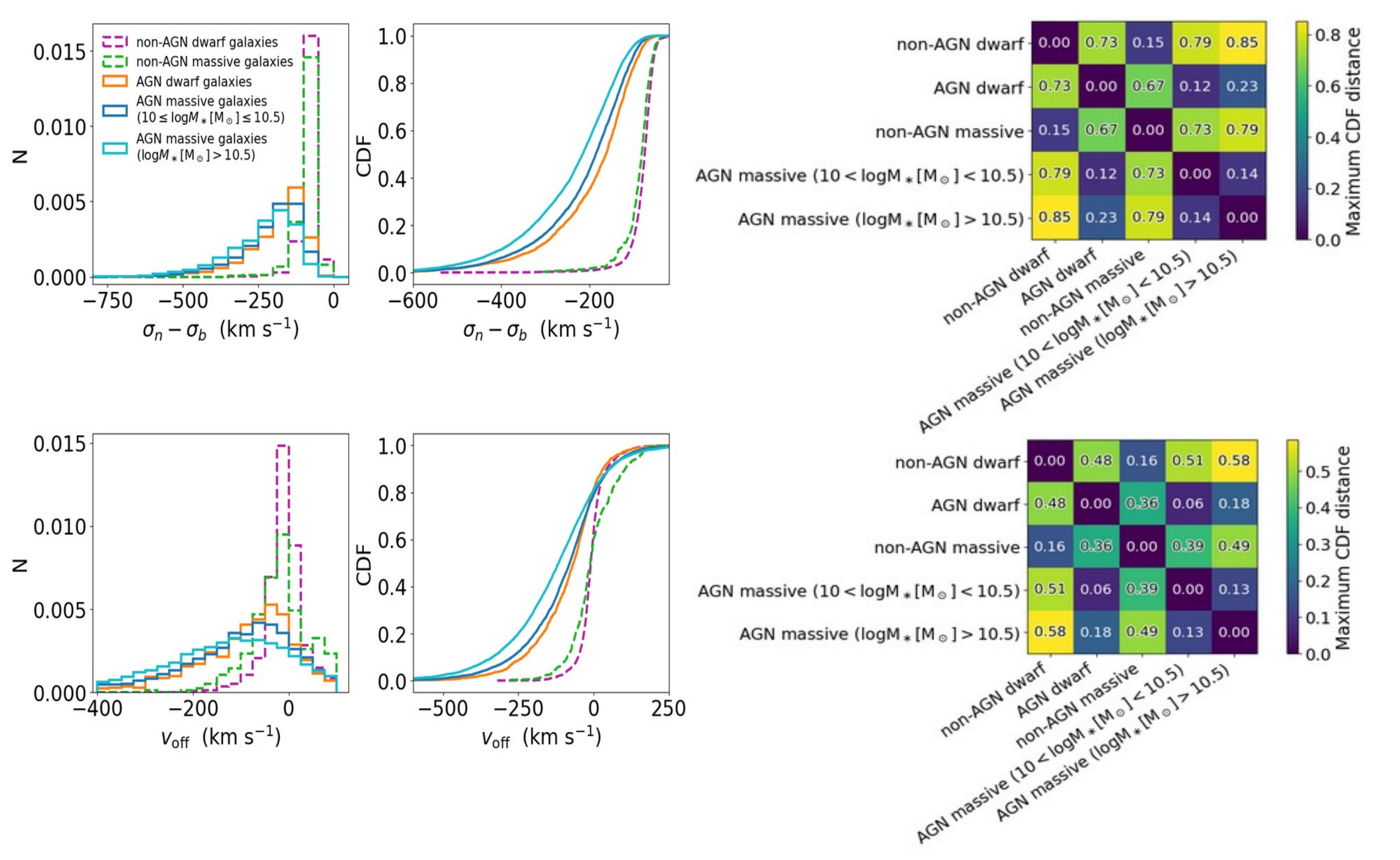}
    \caption{Left: Histograms representing the velocity dispersion difference between the narrow and the broad component of the [OIII]$\lambda5007$ emission line ($\sigma_n-\sigma_b$) (top row) and the velocity offset ($\mathrm{v_{off}}$) between these components (bottom row). Five different distributions are shown: AGN dwarf galaxies (orange line), non-AGN dwarf galaxies (purple dashed line), AGN massive galaxies (split into two stellar-mass bins, blue and cyan lines) and non-AGN massive galaxies (green dashed line). Middle: CDF of the histograms. Right: Heat map of the 5x5 KS statistic matix which represents the maximum distance between the CDFs.}
  	\label{fig:parameters}
\end{figure*}

\subsubsection{Outflow velocity}

There are multiple parameters to define the outflow velocity, but the most widely used is the W$_{80}$ velocity, defined as the velocity width that contains 80$\%$ of the flux of the [OIII] emission line. If the emission line is fitted by a single Gaussian, the W$_{80}$ can be computed as W$_{80}=1.09\times \mathrm{FWHM}$ (Full Width at Half Maximum). As we fit the [OIII] emission line with two components, we performed a non-parametric measurement of the W$_{80}$ velocity (e.g. \citealt{liu2013observations}; \citealt{liu2020integral}; \citealt{rodriguez2025manga}):

\begin{equation}
    \mathrm{W_{80}} = v_{90} - v_{10}
\end{equation}

where $v_{90}$ and $v_{10}$ are velocities at 10th and 90th percentile of the total flux. To calculate the percentiles, it is necessary to compute the cumulative flux distribution ($\phi(v)$) as a function of the velocity. The cumulative flux distribution was evaluated using the best-fitting \texttt{EmFit} model of the [OIII] emission line, and the velocities $v_{90}$ and $v_{10}$ were obtained by numerically inverting the modeled cumulative distribution.

\begin{equation}
    \phi(v)\equiv \int_{-\infty}^{v} F_\mathrm{v}(v')dv'
\end{equation}

where $F_\mathrm{v}$ is the spectral flux density parametrized as a sum of two Gaussian distributions corresponding to the narrow and the broad component of the [OIII] emission line (\citealt{rodriguez2025manga}).

\subsubsection{Energetic properties of the outflows}

The ionized gas mass of the outflows (M$_\mathrm{out}$) can be calculated by using either the [OIII]$\lambda$5007 luminosity or using the Balmer lines H$\alpha$ or H$\beta$ (assuming an intrinsic line ratio H$\alpha$/H$\beta=2.9$), following \cite{osterbrock2006astrophysics} and \cite{carniani2015ionised} equations:

\begin{equation}
\begin{aligned}
    M_{\mathrm{out}} &= 4.48\,M_\odot 
    \left( \frac{L_{H\alpha,\mathrm{corr}}}{10^{35}\ \mathrm{erg\ s^{-1}}} \right)
    \left( \frac{\langle n_e \rangle}{100\ \mathrm{cm^{-3}}} \right)^{-1} \\[1em]
    M_{\mathrm{out}} &= 1.7\times 10^{3}\,
    \frac{m_p\,C\,L_{[\mathrm{O\,III}],\mathrm{corr}}}{
    10^{[\mathrm{O/H}]-[\mathrm{O/H}]_\odot}\,j_{[\mathrm{O\,III}]}\,\langle n_e \rangle}
\end{aligned}
\end{equation}

where $L_\mathrm{H\alpha \ corr}$ is the total luminosity of H$\alpha$ corrected for extinction using \cite{calzetti2000dust} extinction law. The electron density, $n_e$, was computed using \texttt{PyNeb} (\citealt{luridiana2015pyneb}) from the [SII]$\lambda6717/\lambda6731$ line ratio and assuming a typical temperature of the narrow-line region of T$_e=10^4$K. $\mathrm{L_{[OIII]\ corr}}$ is the total luminosity of the [OIII] line corrected from extinction using the \cite{lamastra2009bolometric} relation, $\mathrm{m_p}$ is the proton mass, $C=<n_e>^2/<n_e^2>$, $\mathrm{j_{[OIII]}}$ is the [OIII] line emissivity also calculated with \texttt{PyNeb}, and $\mathrm{{[O/H]-[O/H]_\odot}}$ is the metallicity relative to the solar. To calculate the outflow mass from the [OIII] luminosity, we assumed $C=1$ and the gas having solar metallicities, $\mathrm{[O/H]=[O/H]_\odot}$. \cite{harrison2018agn} remarked the importance of using recombination lines (e.g. H$\alpha$) as they are relatively insensitive to the ionization state and elemental abundances of the outflow. In the catalogue derived from this paper, we reported the outflow mass obtained using both expressions and the corresponding derived properties. 

The mass rate (dM/dt), the energy rate (dE/dt), and the momentum rate (dP/dt) are calculated from the following equations (e.g. \citealt{harrison2014kiloparsec,harrison2018agn}; \citealt{liu2020integral}):

\begin{equation}
\label{eq:mass_rate}
    \mathrm{dM/dt=\frac{M_{out}W_{80}}{R_{out}}}
\end{equation}

\begin{equation}
\label{eq:energy_rate}
    \mathrm{dE/dt=\frac{W_{80}^2+A\sigma_{out}^2}{2}dM/dt}
\end{equation}

\begin{equation}
\label{eq:momentum_rate}
    \mathrm{dP/dt=W_{80} dM/dt}
\end{equation}

We assumed W$_{80}$ as the outflow velocity and A, which is a constant that determines the contribution of the velocity dispersion, being negligible (as as W$_{80}$ already traces the integrated kinematic structure of the emission-line profile). We computed the outflow radius from the FWHM of the Point Spread Function (PSF), $\mathrm{R_{out}}=\frac{1}{2}\times \mathrm{FWHM(PSF)}$ (\citealt{liu2020integral}). The FWHM(PSF) was obtained from the Tractor Catalogue\footnote{https://www.legacysurvey.org/dr9/catalogs/} of the DESI Legacy Imaging Survey (\citealt{dey2019overview}). The Tractor Catalogue gives the FWHM(PSF) for the g-band, r-band and z-band with a Delivered Image Quality (DIQ) of 1.29, 1.18, 1.11 arcsec and depth of 23.95, 23.54 and 22.50 magnitude in AB system. Although the z-band is less affected by seeing, the r-band offers a favourable combination of resolution and depth, this being the selected band when computing the outflow radius. We remark that the FWHM(PSF) represents the angular resolution of the DESI Legacy Imaging Survey, and the PSF-based approach assumes that the outflow is centrally concentrated, and does not account for intrinsic geometry, inclination, or the clumpiness of the outflowing gas. This approach, together with the systematic uncertainties associated with the electron density (typically biased toward lower values due to the small separation between the doublet lines which hampers the isolation of the outflowing contribution) can  introduce large systematic uncertainties of up to an order of magnitude in the derived energetic properties, as discussed in \cite{harrison2018agn}. However, the median outflow radius in our sample ($\mathrm{R_{out}}\sim2.5$ kpc) was consistent with the values reported for spatially resolved AGN outflows in dwarf galaxies, which typically range from $\sim$ 1 kpc to 5 kpc (e.g. \citealt{liu2020integral}; \citealt{rodriguez2025manga}).

\begin{figure*}
\centering
\includegraphics[width=\textwidth]{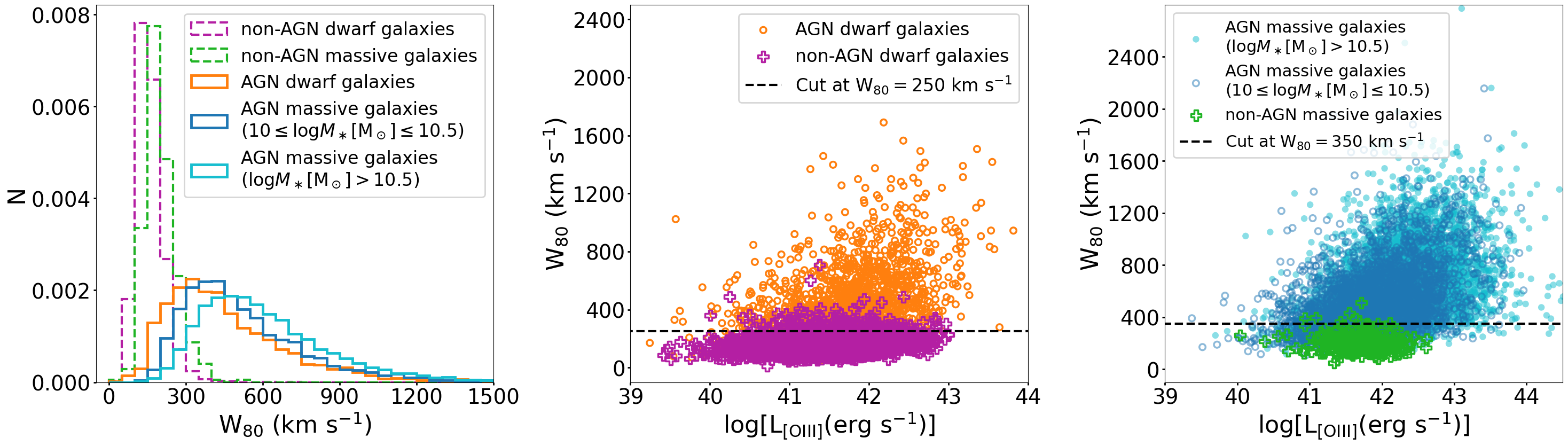} 
\caption{Left: W$_{80}$ outflow velocity distribution for the sample of AGN and non-AGN dwarf and massive galaxies (same color-coded as Fig. \ref{fig:parameters}). Middle and right: W$_{80}$ velocity versus [OIII] luminosity in dwarf and massive galaxies, respectively. The black dashed lines represent the cut on W$_{80}=250$ and 350 km s$^{-1}$ that marks the separation between the distributions of AGN and non-AGN dwarf and massive galaxies, respectively.}
\label{figure:W80}
\end{figure*}

\begin{table*}
\centering
\caption{Exemplary 10 rows of the sample of 1,240 dwarf galaxies with AGN outflows signatures}
\resizebox{\textwidth}{!}{
\begin{tabular}{ccccccccccccccc}
\hline
TARGETID & RA & DEC & z & log($\mathrm{M_\ast}$)& log($\mathrm{L_{[OIII]}}$) & SFR & W$_{80}$ & $\mathrm{R_\mathrm{out}}$ & log($\mathrm{M_\mathrm{out}}$) & log(dM/dt) & log(dE/dt)& log(c dP/dt)\\
& (J2000) & (J2000) &  & (M$_{\odot}$)& erg s$^{-1}$ &(M$_{\odot}$ yr$^{-1}$) & (km s$^{-1}$) & (kpc) & (M$_{\odot}$) & (M$_{\odot}$ yr$^{-1}$) &(erg s$^{-1}$) &(L$_\odot$)  \\
(1) & (2) & (3) & (4) & (5) & (6) & (7)& (8)&(9)&(10)&(11)&(12)&(13) \\ \hline \hline
39627485180069576 & 14.394320 & -12.398983 & 0.4431 & 9.8 $\pm$ 0.2 & 42.2 $\pm$ 0.1 & 3.8$\pm$1.0 & 606 $\pm$ 15 & 5.6 & 6.4 $\pm$ 0.1 & -0.5 $\pm$ 0.1 & 41.15 $\pm$ 0.03 & 9.9 $\pm$ 0.1 \\
39627897459180101 & 242.45089 & 4.58136832 & 0.0644 & 9.7 $\pm$ 0.3  & 41.97 $\pm$ 0.01 & 2.61 $\pm$ 0.05 & 993 $\pm$ 3  & 0.9 & 6.52 $\pm$ 0.01 & 0.57  $\pm$ 0.01 & 42.668  $\pm$ 0.002 & 11.245  $\pm$ 0.009\\
39627678206138177 & 127.822840 & -4.466390 & 0.1293 & 9.79 $\pm$ 0.08 & 41.96 $\pm$ 0.02 & 2.07 $\pm$ 0.09 & 275 $\pm$ 4 & 1.4 & 6.45 $\pm$ 0.02 & -0.25  $\pm$ 0.02 & 40.73  $\pm$ 0.01 & 9.86  $\pm$ 0.02 \\
39628004707533738 & 195.457513 & 8.9023866 & 0.2725 & 9.9 $\pm$ 0.2 & 42.34 $\pm$ 0.06 & 6.45 $\pm$ 0.8 & 695 $\pm$ 14 & 3.2 & 7.81 $\pm$ 0.05 & 1.15  $\pm$ 0.05 & 42.94  $\pm$ 0.01 & 11.67  $\pm$ 0.05 \\
39627754332753885 & 348.375294 & -1.402816 & 0.3423 & 9.9 $\pm$ 0.1  & 41.87 $\pm$ 0.03 & 2.7 $\pm$ 0.2 & 486 $\pm$ 15 & 4.2 & 7.03 $\pm$ 0.03 & 0.11  $\pm$ 0.03 & 41.58  $\pm$ 0.01 & 10.47  $\pm$ 0.03 \\
39627947300096866 & 345.073691 & 6.478247  & 0.0391 & 9.8 $\pm$ 0.1 & 40.43 $\pm$ 0.1 & 0.07 $\pm$ 0.01 & 391 $\pm$ 4 & 0.6 & 5.0 $\pm$ 0.1 & -1.1  $\pm$ 0.1 & 40.16  $\pm$ 0.02 & 9.1  $\pm$ 0.1 \\
39628176778856744 & 262.269628 & 16.27945 & 0.1123 & 9.8 $\pm$ 0.2  & 41.01 $\pm$ 0.05  & 0.41 $\pm$ 0.04 & 745 $\pm$ 24 & 1.3 & 7.11 $\pm$ 0.05 & 0.88  $\pm$ 0.05 & 42.73  $\pm$ 0.01 & 11.43  $\pm$ 0.05 \\
39628244076463040 & 148.333185 & 19.237719 & 0.3812 & 9.6 $\pm$ 0.3 & 41.89 $\pm$ 0.03  &21.9 $\pm$ 1.4 & 661 $\pm$ 14 & 4.6 & 8.12 $\pm$ 0.03 & 1.29  $\pm$ 0.03 & 43.03  $\pm$ 0.01 & 11.78  $\pm$ 0.03 \\
39628497198515802 & 235.843825 & 30.697320 & 0.4262 & 9.9 $\pm$ 0.2  & 42.32 $\pm$ 0.07 & 37.2 $\pm$ 6 & 449 $\pm$ 4 & 7.3 & 8.11 $\pm$ 0.07 & 0.91  $\pm$ 0.07 & 42.31  $\pm$ 0.002 & 11.23  $\pm$ 0.07 \\
39632976539028387 & 227.842378 & 34.819971 & 0.2177 & 9.7 $\pm$ 0.2 & 41.92 $\pm$ 0.09 & 1.9 $\pm$ 0.4 & 524 $\pm$ 5& 3.7 & 6.87 $\pm$ 0.08 & 0.04  $\pm$ 0.08 & 41.58  $\pm$ 0.02 & 10.43  $\pm$ 0.08 \\ \hline 
\end{tabular}}
(1) Unique DESI Target ID; (2,3) Right Ascension and Declination; (4) redshift; (5) galaxy stellar mass from SED fitting; (6) extinction-corrected [OIII]$\lambda 5007$ luminosity of the galaxy; (7) star formation rate derived from extinction-corrected H$\alpha$; (8) outflow W$_{80}$ velocity; (9) radius of the outflow; (10) ionized gas mass of the outflow; (11) ionized gas mass outflow rate; (12) ionized gas kinetic energy outflow rate; (13) ionized gas momentum outflow rate.
\label{tab:summary}
\end{table*}

\section{Results and Discussion}
\label{sect:3}

\subsection{Driving mechanism of the outflows}
\label{sect:driving}

When discussing outflows, it is essential to understand their origin, as the driving mechanism (whether AGN activity or stellar feedback) directly influences their kinematic and energetic properties. In this section, we study the origin of the detected outflows by (i) comparing the properties of the [OIII] emission line between AGN and non-AGN dwarf and massive galaxies with outflow signatures; (ii) examining the differences in the W$_{80}$ outflow velocity between AGN and non-AGN dwarf and massive galaxies with outflow signatures; and (iii) comparing the outflow kinetic energy rate versus the SFR of the AGN dwarf galaxies to determine whether the outflows are compatible with a stellar origin or not.

\subsubsection{[OIII] emission line properties}
\label{sect:param}
We compare two different parameters of the [OIII] emission line that are crucial when studying outflows: the difference in the velocity dispersion between the narrow and the broad component ($\sigma_n-\sigma_b$) and the velocity offset, $\mathrm{v_{off}}$, between the broad and the narrow component (being the broad component blueshifted if $\mathrm{v_{off}}<0$ and redshifted if $\mathrm{v_{off}}>0$, the latter could arise from unresolved biconical outflows affected by projection effects). In Fig. \ref{fig:parameters} two different histograms are shown representing these two parameters, in addition to the Cumulative Distribution Function (CDF) from five different samples: AGN dwarf galaxies, non-AGN dwarf galaxies, AGN massive galaxies (split into two stellar-mass bins) and non-AGN massive galaxies\footnote{We did not split the mass for the non-AGN massive galaxies because the mass range of this sample is contained within the lower-mass bin of the sample of AGN massive galaxies.}. When comparing these normalized histograms and the corresponding CDFs, two different behaviours are observed: one associated to the AGN galaxies and another associated to the non-AGN galaxies, independently of the masses. We do a qualitative inspection of the differences between the four distributions by comparing the Kolmogorov-Smirnov (KS) statistics (i.e. maximum distance between two CDFs). We observe that the AGN galaxies have on average larger ($\sigma_n-\sigma_b$) compared to non-AGN galaxies (with a median of -199 and -75 km s$^{-1}$ for AGN  and non-AGN, respectively), indicating that the outflow components are broader relative to the narrow components in AGN galaxies. We also find that outflows from AGN are, on average, more blueshifted than those in non-AGN outflows, with median velocity shifts of -96 and -9 km s$^{-1}$ for AGN and non-AGN galaxies, respectively. The right panels of Fig. \ref{fig:parameters} show the heat-maps that represent the $5\times5$ KS statistics matrix. The elements of the matrix with lower values represent similarity between the distributions. In both matrices, the lower values are those comparing AGN dwarf and massive galaxies (0.12 and 0.23 for the $\sigma_n-\sigma_b$ distribution,  0.06 and 0.18 for the $\mathrm{v_{off}}$  distribution considering the two stellar-mass bins)  and those comparing non-AGN dwarf and massive galaxies (0.15 for the $\sigma_n-\sigma_b$ distribution and 0.16 for the $\mathrm{v_{off}}$ distribution). These values suggest that the distributions of AGN in dwarf galaxies and massive galaxies are similar, but those between AGN and non-AGN are different. However, there is overlap between the two samples, indicating that some AGN galaxies may exhibit properties similar to those of non-AGN systems, although the AGN population shows, on average, broader and more shifted components. This suggests that AGN activity enhances the likelihood of producing more extreme ionized gas kinematics, and is therefore the most probable driving mechanism of the outflows in these galaxies.

\begin{figure}
\centering
\includegraphics[width=0.48\textwidth]{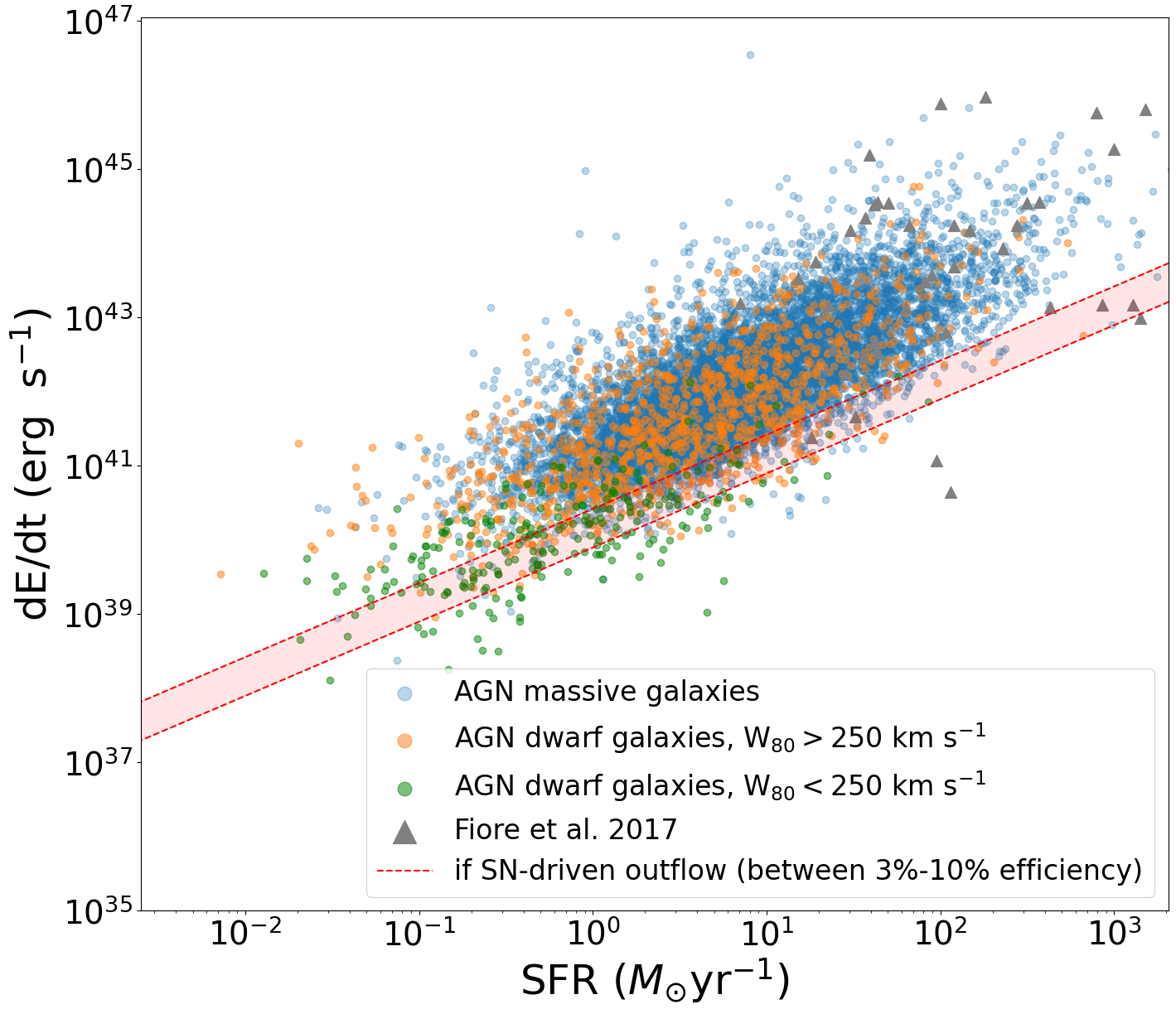} 
\caption{Outflow kinetic energy rate versus SFR of the host galaxies. We differentiate between the sample of dwarf galaxies with AGN outflows with W$_{80}\geq250$ km s$^{-1}$ (orange dots) and those with W$_{80}<250$ km s$^{-1}$ (green dots). We also plot the sample of massive galaxies with AGN outflows (blue dots) and the sample of ionized outflows from \protect\cite{fiore2017agn}. The shaded area in between the red dashed lines represents the expected trend if SNe were the driving mechanism of the outflows considering a SN efficiency between 3$\%$ and 10$\%$.}
\label{figure:SFR}
\end{figure}

\subsubsection{Outflow velocity diagnostic}
\label{velocity_cut}
We study the differences between the W$_{80}$ outflow velocity between the two pairs of samples of dwarf and massive galaxies: AGN and non-AGN with outflow signatures. In the left panel of Fig. \ref{figure:W80} we show five histograms of the W$_{80}$ velocity for the samples of AGN and non-AGN dwarf and massive galaxies with outflow signatures. Two different distributions are observed: while the sample of non-AGN galaxies has lower outflow velocities (median value of 152 and 190 km s$^{-1}$ for dwarf and massive galaxies, respectively), the sample of AGN galaxies presents a Gaussian distribution with a tail towards higher W$_{80}$ values (median value of 405 for dwarf and 465 and 561 km s$^{-1}$ for the two stellar-mass bins of massive galaxies, respectively). This observed difference allows us to propose a new cut to distinguish between AGN outflows and outflows driven by stellar processes. The W$_{80}$ 95th percentile of the non-AGN dwarf galaxies with outflow signatures is 250 km s$^{-1}$. In contrast, 83$\%$ of the AGN dwarf galaxies with outflow signatures have W$_{80}$ velocity greater than 250 km s$^{-1}$. Consequently, we apply a cut in W$_{80}=250$ km s$^{-1}$ to distinguish between outflows ionized by the AGN versus by stellar processes in dwarf galaxies (although there will still be some contamination from outflows driven by stellar processes in the sample with W$_{80}>250$ km s$^{-1}$; \citealt{rubin2010persistence}; \citealt{concas2017light}; \citealt{del2026ga}). In the middle panel of Fig. \ref{figure:W80}, this criterion is well illustrated when comparing the outflow velocity against the [OIII] luminosity in dwarf galaxies. After applying the W$_{80}=250$ km s$^{-1}$ cut to the AGN dwarf galaxies with outflow signatures, we end up with a sample of 1,240 dwarf galaxies with outflows that we associate with the AGN (the properties of 10 of them are listed in Table \ref{tab:summary}). Following the same argument, for the sample of massive galaxies with AGN outflow signatures  we apply a velocity threshold of W$_{80}=$350 km s$^{-1}$ (see right panel of Fig. \ref{figure:W80}), where 98$\%$ of the non-AGN massive galaxies are below this limit, while 75$\%$ and 88$\%$ of AGN massive galaxies in the two-stellar mass bins lie above it, resulting in a sample of 9,080 galaxies. We note that the W$_{80}$ thresholds adopted in this work (250 km s$^{-1}$ for dwarf galaxies and 350 km s$^{-1}$ for massive galaxies) are lower than the typically considered in the literature for massive galaxies, W$_{80}>500$ km s$^{-1}$ (\citealt{wylezalek2020ionized}). Although  our thresholds are less conservative than those used in previous works, they are empirically motivated by the observed W$_{80}$ distributions of AGN and non-AGN galaxies in each stellar-mass regime, and are better suited to capturing those AGN outflows with moderate velocities that would be missed by more stringent cuts. 

\begin{figure*}
\centering
\includegraphics[width=\textwidth]{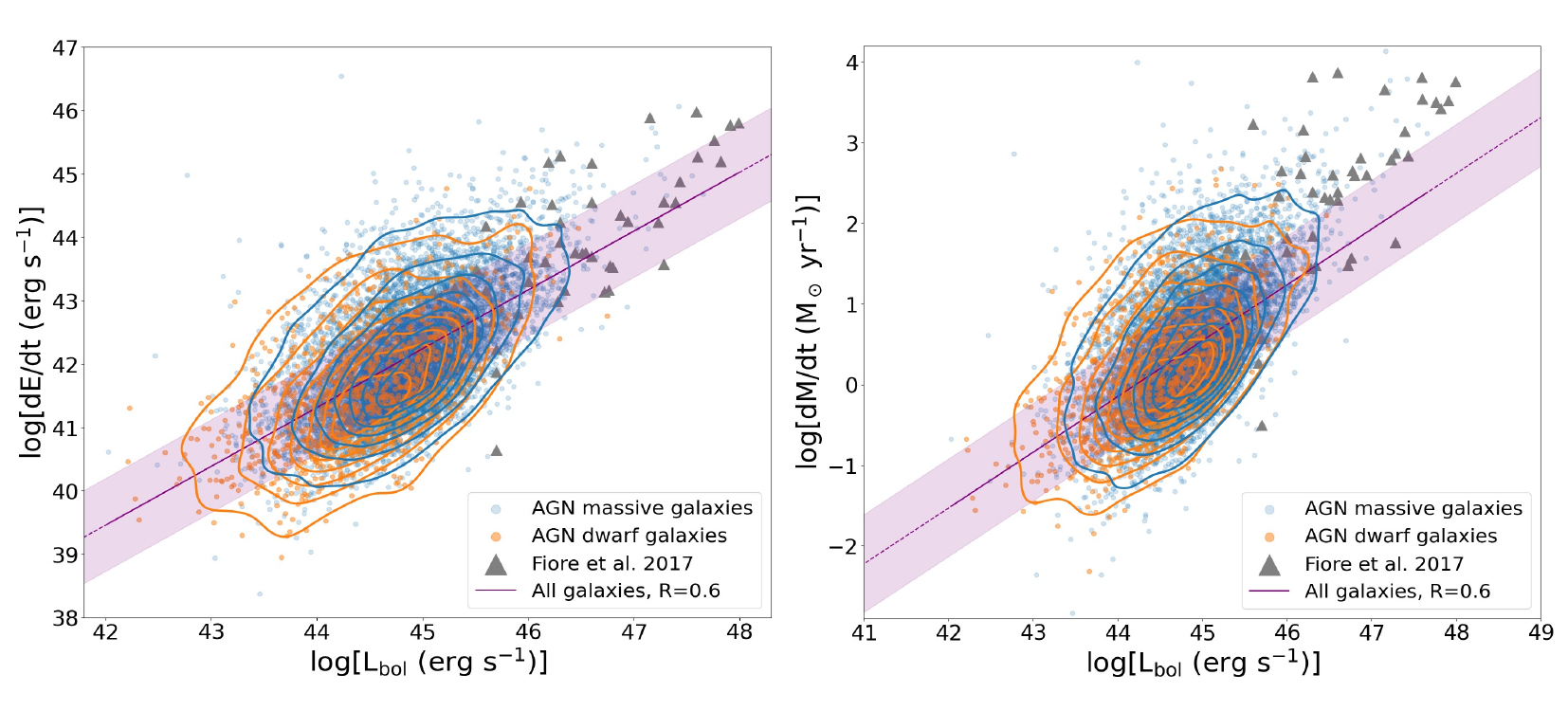} 
\caption{AGN outflow kinetic energy rate (left panel) and mass rate (right panel) versus AGN bolometirc luminosity. The sample of both dwarf and massive galaxies with AGN outflows are shown (orange and blue dots, respectively), in addition to the AGN outflows in massive galaxies from \protect\cite{fiore2017agn}. The purple solid line is a regression fit considering all the galaxies studied in this work. The shaded area corresponds to the 1$\sigma$ error of the regression line. In both regressions, the correlation coefficient is R=0.6.}
\label{figure:scaling}
\end{figure*}

\subsubsection{Kinetic energy rate versus SFR}
\label{kinetic_vs_sfr}

When representing the kinetic energy rate of the outflows against the SFR, a correlation between them is expected if they are driven by SNe (\citealt{fiore2017agn}). In Fig. \ref{figure:SFR}, we show the kinetic energy rate of our sample of 1,502 AGN dwarf galaxies with outflow signatures versus the SFR. The SFR is computed using the equation $\mathrm{log(SFR)}=\mathrm{log(L}_{H\alpha})-41.27$ from \cite{kennicutt2012star}, after correcting L$_{H\alpha}$ for extinction using the \cite{calzetti2000dust} extinction law, computing the total flux of H$\alpha$ and H$\beta$ to obtain the Balmer decrement and the color excess. We point out that L$_{H\alpha}$ is contaminated by AGN contribution and therefore, the SFR may be overestimated\footnote{We did not adopt the SFR estimates from the DESI DR1 VAC of physical properties, as these values may be significantly biased in the low-mass regime. The SED fitting relies on a limited set of broad-band photometric measurements (g, r, z, W1, W2, W3, and W4, when available), which provide only weak constraints on recent star formation. In particular, the lack of UV observations, far-infrared measurements tracing dust-reprocessed star formation, and emission-line information introduces substantial degeneracies between star formation history, dust attenuation, and stellar population properties. In addition, the adopted templates assume solar metallicity, which is unlikely to be representative of many dwarf galaxies. As a result, the inferred SFRs are subject to substantial systematic uncertainties and are systematically higher than those derived from H$\alpha$ measurements.}. We compute the expected relation if SNe were the driving mechanism of the outflows by assuming 0.0083 SN rate per solar mass of newly formed stars (Chabrier initial mass function), a total luminosity for each SN of 10$^{51}$ erg s$^{-1}$, and an efficiency ranging from 3-10$\%$ following \cite{fierlinger2016stellar}. In Fig. \ref{figure:SFR}, this predicted SN-driven outflow scenario is shown as the shaded area between the red dashed lines, with the 3$\%$ efficiency corresponding to the lower boundary and the 10$\%$ efficiency to the upper boundary. In order to be consistent with Sect. \ref{velocity_cut}, we distinguish between the outflows with W$_{80}$ velocity greater than 250 km s$^{-1}$ and those whose velocity is under this limit. Under the 10$\%$ SN efficiency assumption, 91$\%$ of the outflows with W$_{80}<250$ km s$^{-1}$ lie below the SN-driven scenario, whereas only 29$\%$ of the outflows with W$_{80}>250$ km s$^{-1}$ occupy that same region. Adopting a 3$\%$ efficiency reduces these fractions to 54$\%$  and 4$\%$, respectively. These results show that while SNe may drive a substantial fraction of the outflows with W$_{80}<250$ km s$^{-1}$, it is energetically insufficient to explain the majority of those with W$_{80}>250$ km s$^{-1}$, the latter being more consistent with an AGN origin, reinforcing the results of Sect. \ref{velocity_cut}.

We have also computed the kinetic energy rate expected from starbust-driven winds including contributions from both stellar winds and SNe. We use the expression $\mathrm{(dE/dt)_{\ast}=7\times 10^{41}(SFR/M_{\odot}yr^{-1})}$ from \cite{veilleux2005galactic}, obtained from models that synthesize the evolution of populations of massive stars using \texttt{Starbust99}\footnote{https://massivestars.stsci.edu/starburst99/docs/default.htm} (\citealt{leitherer1999starburst99}). For our sample, we obtain a median value $\mathrm{(dE/dt)_\ast\sim~10^{42.4}}$, that is approximately one order of magnitude higher than the median value of the kinetic energy rate obtained using Eq. \ref{eq:energy_rate}. Consequently, the starbust-driven winds could explain these outflows. However, as noted above, the results obtain from starbust winds are overestimated due to the AGN contribution when computing the SFR.

In the end, our final sample of dwarf and massive galaxies with AGN outflows consists of 1,240 and 9,080 objects with W$_{80}$ velocity greater than 250 km s$^{-1}$ and 350 km s$^{-1}$, respectively.

\subsection{Scaling relations: comparison with massive galaxies}

In this section, we compare the energetic properties of the AGN outflows between dwarf and massive galaxies. In the left panel of Fig. \ref{figure:scaling} we show the kinetic energy rate versus the bolometric luminosity, L$_\mathrm{bol}$, for the sample of dwarf and massive galaxies with AGN outflows studied in this work, as well as the massive galaxies with AGN outflows studied in \cite{fiore2017agn}. The data from \cite{fiore2017agn} includes galaxies with outflows traced by [OIII], H$\alpha$, and/or H$\beta$ lines, covering from low-redshift systems at z$<$0.2 to luminous/hyper-luminous qusars at z$=$2-3, the latter representing the data with highest bolometric luminosities. We compute the bolometric luminosity from the L$_\mathrm{[OIII]}$ following the \cite{lamastra2009bolometric} relation. In the right panel of Fig. \ref{figure:scaling} we represent the outflow mass rate versus L$_\mathrm{bol}$ for the same sample of dwarf and massive galaxies with AGN outflows in addition to those from \cite{fiore2017agn}. We observe that a linear relationship spans in both panels (purple lines of Fig. \ref{figure:scaling} with a Pearson correlation coefficient R = 0.6 and p-value $\ll 0.01$), without a clear distinction between dwarf and massive galaxies, with overlapping distributions, reinforcing the results discussed in Sect. \ref{sect:param}. Although dwarf galaxies occupy the regions with lower luminosity, we observe through the plotted contours that a significant part of them can be as energetic as massive galaxies. These results provide new insights into the behaviour and properties of AGN outflows in the low-mass regime, which appear to follow the same linear trend observed in massive galaxies. This is in agreement with the results found in \cite{rodriguez2025manga}, who studied a sample of 13 dwarf galaxies with AGN outflows suggesting a linear relation between the kinetic energy rate and the bolometric luminosity throughout the entire mass regime. These linear relationships indicate that AGN outflows in dwarf galaxies appear to be as energetic as those in massive galaxies in the same bolometric luminosity range.

\subsection{Can AGN outflows impact the host star formation?}
In this section, we investigate whether the AGN outflows in dwarf galaxies can regulate star formation in their host galaxies by studying the escape velocity of the host dark matter halo and the coupling efficiency of the outflows. We also study the incidence of the AGN outflows in the dwarf galaxy population.

\subsubsection{Escape velocity}

\begin{figure}
\centering
\includegraphics[width=0.45\textwidth]{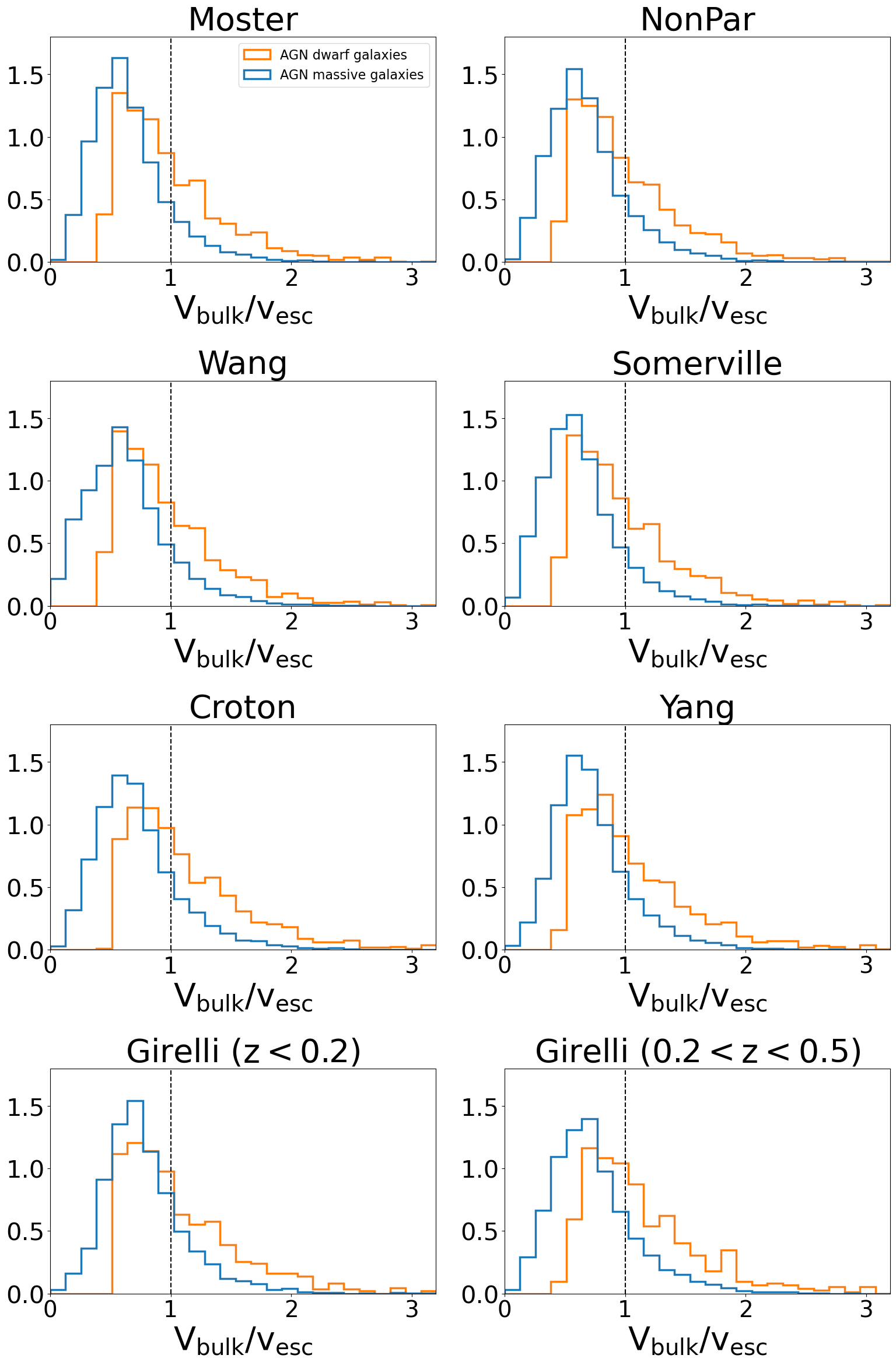} 
\caption{Distribution of the ratio between the V$_\mathrm{bulk}$ velocity and the escape velocity, v$_\mathrm{esc}$, necessary to escape the gravitational potential produced by the dark matter halo. We show the distribution for eight different models used to compute the dark matter halo masses: the \protect\cite{moster2010constraints, moster2013galactic}, the non-parametric (\protect\citealt{vale2006non}; \protect\citealt{conroy2006modeling}; \protect\citealt{shankar2006new}), the \protect\cite{wang2006modelling}, the \protect\cite{croton2006many}, the \protect\cite{yang2008galaxy}, the \protect\cite{somerville2008semi} and the \protect\cite{girelli2020stellar} model, differentiating between those galaxy with z $<0.2$ and 0.2 $\leq$ z $<$ 0.5 in the latest. Those galaxies with V$_\mathrm{bulk}/\mathrm{v}_\mathrm{esc}>1$ (represented by a vertical black dashed line) are capable to escape the host dark matter halo.}
\label{figure:vesc}
\end{figure}

We study if the AGN outflows are able to escape from the gravitational potential produced by the dark matter halo of the host galaxy. For this purpose, we use the abundance matching method that relates the stellar mass of the galaxy with the dark matter halo mass ($M_h$) assuming a Navarro-Frenk-White profile (NFW, \citealt{navarro1997universal}):

\begin{equation}
\frac{M_{\ast}}{M_h}=2\left(\frac{M_\ast}{M_h}\right)_0\left[\left(\frac{M_h}{M_1}\right)^{-\beta}+\left(\frac{M_h}{M_1}\right)^\gamma\right]^{-1}
\end{equation}

where $(M_\ast/M_h)_0$, $M_1$, $\beta$, $\gamma$ are four parameters that depend on the model used (see Table \ref{tab:parameters}). We compute the different halo masses based on eight different models of the stellar-to-halo mass relation, each corresponding to a different halo occupation model: the \cite{moster2010constraints, moster2013galactic} model, the non-parametric model (\citealt{vale2006non}; \citealt{conroy2006modeling}; \citealt{shankar2006new}), the \cite{wang2006modelling} model, the \cite{croton2006many} model, the \cite{yang2008galaxy} model, the \cite{somerville2008semi} and the \cite{girelli2020stellar} model. For the latest model, we use the redshift-dependent parameters for the two gaps that fall within our range: z $<0.2$ and $0.2\leq \mathrm{z} < 0.5$. Then, we can compute the escape velocity, $v_\mathrm{esc}$ using the equation defined in \cite{manzano2019agn}:

\begin{equation}
    v_\mathrm{esc}^2=2|\phi(r=0)|
\end{equation}

where $\phi(r=0)=-4/3 \ cg(c)\pi Gr_v^2v\rho_c^0$ is the gravitational potential from a spherical NFW profile, being $c=10$ the concentration parameter, $g(c)=1/\mathrm{ln}(1+c)-c/(1+c)$ a function of the concentration parameter, $v=200$ the virial overdensity, $\rho_c^0=277.5M_\odot\mathrm{kpc}^{-3}$ the present critical density, and $r_v=(3M_h/4\pi v\rho_c^0)^{1/3}$ the virial radius. 

\begin{table}
    \caption{Stellar-to-halo mass relation model parameters}

    \begin{tabular}{ccccccccccccccc}
    \hline
    Model & $(M_\ast / M_h)_0$ & log$M_1$ & $\beta$ & $\gamma$ \\ \hline \hline
    \cite{moster2010constraints} & 0.0282 & 11.844 & 1.06 & 0.556 \\
    \cite{vale2006non} & 0.0324 & 11.766 & 1.43 & 0.565 \\
    \cite{wang2006modelling} & 0.0319 & 11.845 & 1.42 & 0.710 \\
    \cite{somerville2008semi} & 0.0276 & 11.888 & 0.98 & 0.629 \\
    \cite{croton2006many} & 0.0405 & 11.742 & 0.92 & 0.610 \\
    \cite{yang2008galaxy} & 0.0384 & 12.067 & 0.71 & 0.698 \\
    \cite{girelli2020stellar}\\ (z $<0.2$) & 0.0465 & 11.770 & 1.00 & 0.702 \\
    \cite{girelli2020stellar}\\ ($0.2\leq \mathrm{z} < 0.5$) & 0.0431 & 11.860 & 0.97 & 0.644 \\ \hline 
    \end{tabular}
	\label{tab:parameters}
\end{table}

In Fig. \ref{figure:vesc}, we show different histograms of the ratio between the bulk-outflow velocity V$_\mathrm{bulk}=\mathrm{W}_{80}/1.3$ (suitable for the spherically symmetric or wide-angle bi-cone outflows, see \citealt{harrison2012energetic, harrison2014kiloparsec}; \citealt{liu2013observations}) and the escape velocity for the two samples of dwarf and massive galaxies with AGN outflows. Those galaxies having a ratio V$_\mathrm{bulk}/v_\mathrm{esc}>1$ are capable of escaping the gravitational potential well produced by the dark matter halo of the host galaxy. For these galaxies, the outflowing gas could impact the star formation of the galaxies as they are more capable of removing gas from the central regions and heating the circumgalactic medium (CGM), thus suppressing the star formation (\citealt{smethurst2021kiloparsec}; \citealt{ivey2026exploring}). The outflowing gas also produces a metal enrichment in the CGM (\citealt{liu2020integral}; \citealt{choi2020impact}; \citealt{martin2024agn}), but as the ejected gas will eventually be unbound from the dark matter halo reaching the intergalactic medium, it will not be cooled and reaccrete onto the galaxy (\citealt{manzano2019agn}), making positive feedback from recycled gas a more unlikely scenario. However, we note that a fraction of the gas may remain bound to the halo.

We perform a two-sample KS-test on the V$_\mathrm{bulk}/v_\mathrm{esc}$ distributions of AGN dwarf and massive galaxies to assess whether they are drawn from the same parent distribution. We reject the null hypothesis (with p-value$\ll0.05$ and KS statistics ranging from 0.3 to 0.4 across the different models). For all the models considered, we find that roughly 30$\%$ more AGN outflows in dwarf galaxies are able to escape the gravitational potential of their dark matter halos than in massive galaxies. This is in agreement with other studies that compare the ratio of the outflow velocity to the escape velocity as a function of the stellar mass, indicating that ionized outflows in dwarf galaxies are more likely to escape from their hosts (\citealt{arribas2014ionized}; \citealt{rodriguez2019properties}; \citealt{schroetter2019muse}; \citealt{xu2022empress}). This result suggests that AGN feedback in dwarf galaxies can be as significant, or even more so, than the AGN feedback in massive galaxies.  

\subsubsection{Coupling efficiency and mass-loading factor}

\begin{figure}
\centering
\includegraphics[width=0.45\textwidth]{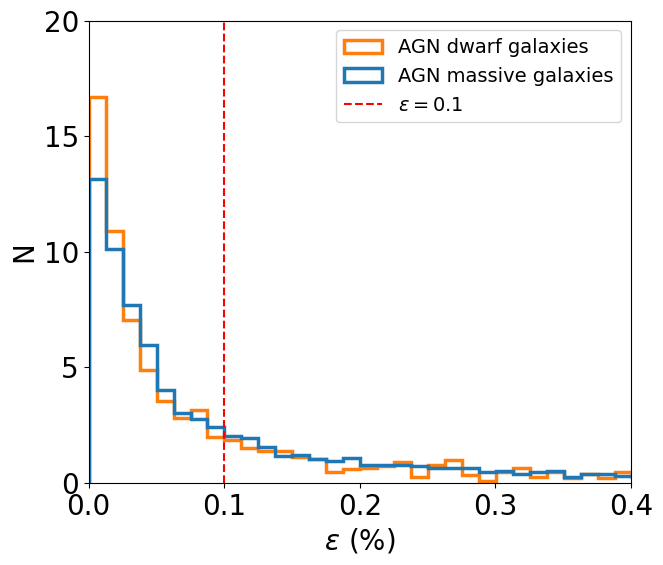}
\caption{Distribution of the coupling efficiency, $\epsilon=$ (dE/dt)$/$L$_\mathrm{bol}$, of the two samples of dwarf and massive galaxies with AGN outflows (orange and blue line, respectively). The vertical red dashed line at $\epsilon=0.001$ (0.1$\%$) shows the minimum coupling efficiency necessary for the outflow to impact the host galaxy.}
\label{efficiency}
\end{figure}

The kinetic coupling efficiency, $\epsilon=(\mathrm{dE/dt})/L_\mathrm{bol}$, is often used to study if the outflows are powerful enough to have an impact on the galaxy. Values above $\epsilon=0.01$ ($1\%$) have been commonly taken as evidence of AGN feedback being able to quench the star formation in massive galaxies (\citealt{harrison2014kiloparsec}; \citealt{feruglio2015multi}; \citealt{rose2018quantifying}; \citealt{costa2018driving}). However, theoretical predictions postulate a coupling efficiency $\epsilon\sim 10^{-3}$ (0.1$\%$) as the minimum value required for outflows to have a significant impact in the host galaxy (\citealt{choi2012radiative}; \citealt{harrison2018agn}; \citealt{vayner2024first}). In Fig. \ref{efficiency}, we show the coupling efficiency distribution for the two samples of massive and dwarf galaxies with AGN outflows. 40$\%$ of dwarf and massive galaxies with AGN outflows, have a coupling efficiency above 0.1$\%$, suggesting that these AGN outflows are potentially capable of having an impact on the evolution of the host galaxy. However, special caution must be taken: first, because the observed coupling efficiencies should not be compared to the simulated one, as they are calibrated quantities rather than predictions; and second, because we are missing parts of the puzzle due to the multi-phase nature of the outflows, with hot ionized gas, neutral atomic gas, and molecular gas outflows contributing to the overall kinetic coupling efficiencies, being the hot ionized gas (T$>10^{6}$K) the one that contributes the most (\citealt{ward2024agn}). Furthermore, it is important to remark that the impact of these outflows is likely to be delayed over long timescales, long after the AGN or the outflow are observed (\citealt{harrison2017impact}; \citealt{angles2017gravitational}; \citealt{costa2018driving}). This represents one of the main challenges in studying AGN feedback, as constraining the long-term impact of the observed outflows, and determining whether they lead to positive or negative feedback, is extremely difficult due to the time-scales and observational uncertainties involved.

The mass-loading factor, $\eta=\mathrm{(dM/dt)/SFR}$, measures how efficiently the outflow is expelling mass relative to the rate at which the galaxy is forming stars. For our sample of dwarf galaxies with AGN outflows, we find a median mass-loading factor of $\sim 0.4$. We consider this value to be a lower limit due to the possible overestimation of the SFR, as discussed in Sect. \ref{kinetic_vs_sfr}. We find that our results are comparable to those reported in previous studies (e.g. \citealt{liu2020integral}; \citealt{ivey2026exploring}), which typically find values of $\eta\sim 0.3$. In our sample, $\sim 20\%$ of the galaxies exhibit $\eta\geq 1$. Consequently, a significant impact on the host galaxies may be expected, as the outflow would be capable of removing gas faster than it is consumed by star formation. However, several important caveats must be considered. Firstly, we note the phase dependence (as discussed previously for the coupling efficiencies), since the mass-loading factors derived from the ionised gas phase alone may underestimate the total mass-loading factor if a substantial fraction of the outflowing mass resides in other gas phases (e.g. \citealt{rupke2017quasar}). Secondly, the mass-loading factor is an instantaneous quantity that can vary on timescales of a few Myr (\citealt{qiu2021mass}). Therefore, it may not be representative of the long-term impact of AGN feedback. Nevertheless, the cumulative effect of multiple AGN episodes could produce the long-term effects in the global properties of the host galaxies (\citealt{piotrowska2022quenching}; \citealt{bluck2022quenching}; \citealt{harrison2024observational}).

\subsubsection{Incidence of AGN outflows in dwarf galaxies}

Comparing the parent sample of low-redshift (z<0.45) dwarf galaxies in DESI DR1 ($\sim$ 2.4M objects) with our final sample of dwarf galaxies hosting AGN outflows (1,240 objects), we find that only $\sim 0.05\%$ of dwarf galaxies exhibit AGN outflows. Since some double-peaked systems may also host AGN outflows, and we excluded them from our sample, this value should be regarded as a lower-limit of the fraction of AGN outflows in dwarf galaxies (see Sect. \ref{sect: double-peaked}). This indicates that such systems are relatively rare within the overall dwarf galaxy population. This low incidence is broadly consistent with previous studies based on integrated spectroscopy, which report AGN fractions below 1$\%$ in low-mass galaxies. Recently, \cite{pucha2026new} find an AGN fraction of $\sim 0.3\%$ using the [NII]-BPT diagram alone (excluding galaxies classified as Composite), while other studies report similarly low values (e.g. \citealt{reines2013dwarf}; \citealt{salehirad2022hundreds}). However, AGN fractions can increase substantially when integral-field spectroscopy is used, reaching values of up to $\sim 20\%$ (e.g. \citealt{polimera2022resolve}; \citealt{mezcua2024manga}). This suggests that AGN activity, and therefore AGN outflows,  in dwarf galaxies may be significantly underestimated by analyses based solely on integrated spectra.

Although the observed fraction of dwarf galaxies hosting AGN outflows is small, the incidence of these systems is a key factor in assessing the overall importance of AGN feedback in dwarf galaxy evolution. Determining its cumulative impact requires not only measuring how frequently AGN outflows occur, but also understanding their duty cycle and the timescales over which the expelled gas can cool, reaccrete, or escape. Therefore, while our results show that AGN outflows could affect individual dwarf galaxies, their role in shaping the dwarf galaxy population as a whole remains uncertain.

\section{Conclusions}
\label{sect:4}
In the last decade, the detection of thousands of AGN in dwarf galaxies, together with the recent studies showing AGN outflows and feedback in these systems, has displaced SNe as the sole mechanism capable of regulating SF in these galaxies. In this paper, we present a sample of dwarf galaxies with AGN outflows identified through the detection of a broad component in the [OIII]$\lambda5007$ emission line. The galaxies are drawn from DESI DR1, and their spectra have been fitted using the \texttt{EmFit} python code. Three additional samples are analyzed to compare their properties: one consisting of massive galaxies with AGN outflows, and the other two consisting of non-AGN dwarf and massive galaxies with outflows. The main results of this work are:

\begin{itemize}
\item{We identify 1,240 dwarf (M$_\ast<10^{10}$M$_\odot$) galaxies with AGN outflow signatures, the largest such sample to date}\\

\item{We find similar distributions for the velocity offset between the narrow and the broad component and for the difference between the velocity dispersions of the narrow and the broad component when comparing matched pairs: AGN dwarf galaxies with AGN massive galaxies, and non-AGN dwarf galaxies with non-AGN massive galaxies. This suggests that the driving mechanisms behind the existence of a broad component in the [OIII] emission line are different when comparing AGN and non-AGN galaxies, independently of the stellar mass of the galaxy.} \\

\item{We find a new cut in the outflow velocity, W$_{80}=250$ for dwarf galaxies and 350 km s$^{-1}$ for massive galaxies, to distinguish between outflows with a stellar origin and outflows with an AGN origin. This result is consistent when comparing the kinetic energy rate versus the SFR, for which those dwarf galaxies with outflow velocities below 250 km s$^{-1}$ are consistent with being driven by SNe.}\\

\item{We find a linear relation spanning the entire mass regime, indicating that AGN outflows in dwarf galaxies behave similarly to those in massive galaxies, albeit at lower luminosities, and thus being a scaled-down version of those in massive galaxies. We point out that, in several cases, we observe that AGN outflows in dwarfs can reach energies comparable to those in massive galaxies.}\\

\item{We find that AGN outflows in dwarf galaxies are more capable of escaping the gravitational potential produced by the dark matter halo than those in massive galaxies. This may indicate that AGN outflows in dwarf galaxies could have a major impact on the host galaxies, as they are more likely to redistribute the gas from the inner regions of the galaxies to the outer regions.}\\

In a future work, we plan to study the presence of AGN outflows in those dwarf  galaxies with redshift z $>0.45$ observed in DESI DR1. DESI has the potential to transform our understanding of galaxy evolution: with its DR1, we have been able to identify more than a thousand AGN outflows in dwarf galaxies greatly increasing the sample available in the literature. Looking ahead, future integral field spectroscopy observations will enable a more detailed characterization of AGN feedback by examining the stellar population properties of these systems, providing new insights into how AGN activity influences their star formation histories and evolution.
 
\end{itemize}

\section*{Acknowledgements}

This work was partially supported by the program Unidad de Excelencia María de Maeztu CEX2020-001058-M. V.R.M. and M.M. acknowledge support from the Spanish Ministry of Science and Innovation through the project PID2024-159201NB-C22 and PRE2022-104649, respectively. R.P. is currently supported by the University of Utah and was also previously supported by the University of Arizona, and in part by NSF NOIRLab. The work of S.J. is supported by NSF NOIRLab, which is managed by the Association of Universities for Research in Astronomy (AURA) under a cooperative agreement with the National Science Foundation. M.S. acknowledges support by the State Research Agency of the Spanish Ministry of Science and Innovation under the grants ’Galaxy Evolution with Artificial Intelligence’ (PGC2018-100852-A-I00) and ’BASALT’ (PID2021-126838NB-I00) and the Polish National Agency for Academic Exchange (Bekker grant BPN/BEK/2021/1/00298/DEC/1). This work was partially supported by the European Union’s Horizon 2020 Research and Innovation program under the Maria Sklodowska-Curie grant agreement (No. 754510). SP is supported by the international Gemini Observatory, a program of NSF NOIRLab, which is managed by the Association of Universities for Research in Astronomy (AURA) under a cooperative agreement with the U.S. National Science Foundation, on behalf of the Gemini partnership of Argentina, Brazil, Canada, Chile, the Republic of Korea, and the United States of America. \\

This material is based upon work supported by the U.S. Department of Energy (DOE), Office of Science, Office of High-Energy Physics, under Contract No. DE–AC02–05CH11231, and by the National Energy Research Scientific Computing Center, a DOE Office of Science User Facility under the same contract. Additional support for DESI was provided by the U.S. National Science Foundation (NSF), Division of Astronomical Sciences under Contract No. AST-0950945 to the NSF’s National Optical-Infrared Astronomy Research Laboratory; the Science and Technology Facilities Council of the United Kingdom; the Gordon and Betty Moore Foundation; the Heising-Simons Foundation; the French Alternative Energies and Atomic Energy Commission (CEA); the National Council of Humanities, Science and Technology of Mexico (CONAHCYT); the Ministry of Science, Innovation and Universities of Spain (MICIU/AEI/10.13039/501100011033), and by the DESI Member Institutions: \url{https://www.desi.lbl.gov/collaborating-institutions}. Any opinions, findings, and conclusions or recommendations expressed in this material are those of the author(s) and do not necessarily reflect the views of the U. S. National Science Foundation, the U. S. Department of Energy, or any of the listed funding agencies.\\

The authors are honored to be permitted to conduct scientific research on I'oligam Du'ag (Kitt Peak), a mountain with particular significance to the Tohono O’odham Nation.\\

Este trabajo forma parte de la tesis doctoral del primer autor, V.R.M. , en el marco del Programa de Doctorado en Física de la Universidad Autónoma de Barcelona.

\section*{Data availability}
The data underlying all figures in the paper will be made available on Zenodo. Additionally, the derived kinematic and energetic properties of the AGN outflows in dwarf galaxies will be provided in two separate tables: one containing the properties derived from the H$\alpha$ emission line and the other containing those derived from the [OIII] emission line.

\bibliographystyle{mnras} % style aa.bst
% \bibliography{Referencias} % your references Yourfile.bib

\begin{thebibliography}{}
\makeatletter
\relax
\def\mn@urlcharsother{\let\do\@makeother \do\$\do\&\do\#\do\^\do\_\do\%\do\~}
\def\mn@doi{\begingroup\mn@urlcharsother \@ifnextchar [ {\mn@doi@} {\mn@doi@[]}}
\def\mn@doi@[#1]#2{\def\@tempa{#1}\ifx\@tempa\@empty \href {http://dx.doi.org/#2} {doi:#2}\else \href {http://dx.doi.org/#2} {#1}\fi \endgroup}
\def\mn@eprint#1#2{\mn@eprint@#1:#2::\@nil}
\def\mn@eprint@arXiv#1{\href {http://arxiv.org/abs/#1} {{\tt arXiv:#1}}}
\def\mn@eprint@dblp#1{\href {http://dblp.uni-trier.de/rec/bibtex/#1.xml} {dblp:#1}}
\def\mn@eprint@#1:#2:#3:#4\@nil{\def\@tempa {#1}\def\@tempb {#2}\def\@tempc {#3}\ifx \@tempc \@empty \let \@tempc \@tempb \let \@tempb \@tempa \fi \ifx \@tempb \@empty \def\@tempb {arXiv}\fi \@ifundefined {mn@eprint@\@tempb}{\@tempb:\@tempc}{\expandafter \expandafter \csname mn@eprint@\@tempb\endcsname \expandafter{\@tempc}}}

\bibitem[\protect\citeauthoryear{Adame et~al.,}{Adame et~al.}{2025}]{adame2025desi}
Adame A.,  et~al., 2025, JCAP, 2025, 028

\bibitem[\protect\citeauthoryear{Anand et~al.,}{Anand et~al.}{2024}]{anand2024archetype}
Anand A.,  et~al., 2024, AJ, 168, 124

\bibitem[\protect\citeauthoryear{Angl{\'e}s-Alc{\'a}zar, Dav{\'e}, Faucher-Gigu{\`e}re, {\"O}zel  \& Hopkins}{Angl{\'e}s-Alc{\'a}zar et~al.}{2017}]{angles2017gravitational}
Angl{\'e}s-Alc{\'a}zar D.,  Dav{\'e} R.,  Faucher-Gigu{\`e}re C.-A.,  {\"O}zel F.,   Hopkins P.~F.,  2017, MNRAS, 464, 2840

\bibitem[\protect\citeauthoryear{Arjona-Galvez, Di~Cintio  \& Grand}{Arjona-Galvez et~al.}{2024}]{arjona2024role}
Arjona-Galvez E.,  Di~Cintio A.,   Grand R.~J.,  2024, \mn@doi [A\&A] {10.1051/0004-6361/202449439}, 690, A286

\bibitem[\protect\citeauthoryear{Arribas, Colina, Bellocchi, Maiolino  \& Villar-Martin}{Arribas et~al.}{2014}]{arribas2014ionized}
Arribas S.,  Colina L.,  Bellocchi E.,  Maiolino R.,   Villar-Martin M.,  2014, A\&A, 568, A14

\bibitem[\protect\citeauthoryear{Assef, Stern, Noirot, Jun, Cutri  \& Eisenhardt}{Assef et~al.}{2018}]{assef2018wise}
Assef R.,  Stern D.,  Noirot G.,  Jun H.,  Cutri R.,   Eisenhardt P.,  2018, ApJS, 234, 23

\bibitem[\protect\citeauthoryear{Baldassare, Geha  \& Greene}{Baldassare et~al.}{2018}]{baldassare2018identifying}
Baldassare V.~F.,  Geha M.,   Greene J.,  2018, ApJ, 868, 152

\bibitem[\protect\citeauthoryear{Baldassare, Geha  \& Greene}{Baldassare et~al.}{2020}]{baldassare2020search}
Baldassare V.~F.,  Geha M.,   Greene J.,  2020, ApJ, 896, 10

\bibitem[\protect\citeauthoryear{Baldwin, Phillips  \& Terlevich}{Baldwin et~al.}{1981}]{baldwin1981classification}
Baldwin J.~A.,  Phillips M.~M.,   Terlevich R.,  1981, PASP, 93, 5

\bibitem[\protect\citeauthoryear{Barai \& de Gouveia Dal~Pino}{Barai \& de~Gouveia Dal~Pino}{2019}]{barai2019intermediate}
Barai P.,  de Gouveia Dal~Pino E.~M.,  2019, MNRAS, 487, 5549

\bibitem[\protect\citeauthoryear{Bichang’a, Kaviraj, Lazar, Jackson, Das, Smith, Watkins  \& Martin}{Bichang’a et~al.}{2024}]{bichang2024}
Bichang’a B.,  Kaviraj S.,  Lazar I.,  Jackson R.~A.,  Das S.,  Smith D. J.~B.,  Watkins A.~E.,   Martin G.,  2024, \mn@doi [MNRAS] {10.1093/mnras/stae1441}, 532, 613

\bibitem[\protect\citeauthoryear{Birchall, Watson  \& Aird}{Birchall et~al.}{2020}]{birchall2020x}
Birchall K.~L.,  Watson M.,   Aird J.,  2020, MNRAS, 492, 2268

\bibitem[\protect\citeauthoryear{Bluck, Maiolino, Brownson, Conselice, Ellison, Piotrowska  \& Thorp}{Bluck et~al.}{2022}]{bluck2022quenching}
Bluck A.~F.,  Maiolino R.,  Brownson S.,  Conselice C.~J.,  Ellison S.~L.,  Piotrowska J.~M.,   Thorp M.~D.,  2022, A\&A, 659, A160

\bibitem[\protect\citeauthoryear{Boquien, Burgarella, Roehlly, Buat, Ciesla, Corre, Inoue  \& Salas}{Boquien et~al.}{2019}]{boquien2019cigale}
Boquien M.,  Burgarella D.,  Roehlly Y.,  Buat V.,  Ciesla L.,  Corre D.,  Inoue A.,   Salas H.,  2019, A\&A, 622, A103

\bibitem[\protect\citeauthoryear{Brodzeller et~al.,}{Brodzeller et~al.}{2023}]{brodzeller2023performance}
Brodzeller A.,  et~al., 2023, AJ, 166, 66

\bibitem[\protect\citeauthoryear{Bruzual \& Charlot}{Bruzual \& Charlot}{2003}]{bruzual2003stellar}
Bruzual G.,  Charlot S.,  2003, MNRAS, 344, 1000

\bibitem[\protect\citeauthoryear{Burke et~al.,}{Burke et~al.}{2022}]{burke2022dwarf}
Burke C.~J.,  et~al., 2022, MNRAS, 516, 2736

\bibitem[\protect\citeauthoryear{Calzetti, Armus, Bohlin, Kinney, Koornneef  \& Storchi-Bergmann}{Calzetti et~al.}{2000}]{calzetti2000dust}
Calzetti D.,  Armus L.,  Bohlin R.~C.,  Kinney A.~L.,  Koornneef J.,   Storchi-Bergmann T.,  2000, ApJ, 533, 682

\bibitem[\protect\citeauthoryear{Capelo, Feruglio, Hickox  \& Tombesi}{Capelo et~al.}{2024}]{capelo2024black}
Capelo P.~R.,  Feruglio C.,  Hickox R.~C.,   Tombesi F.,  2024, in , Handbook of X-ray and Gamma-ray Astrophysics.
Springer, pp 4567--4616

\bibitem[\protect\citeauthoryear{Carniani et~al.,}{Carniani et~al.}{2015}]{carniani2015ionised}
Carniani S.,  et~al., 2015, A\&A, 580, A102

\bibitem[\protect\citeauthoryear{Chabrier}{Chabrier}{2003}]{chabrier2003galactic}
Chabrier G.,  2003, ApJ, 586, L133

\bibitem[\protect\citeauthoryear{Chaussidon et~al.,}{Chaussidon et~al.}{2023}]{chaussidon2023target}
Chaussidon E.,  et~al., 2023, ApJ, 944, 107

\bibitem[\protect\citeauthoryear{Choi, Ostriker, Naab  \& Johansson}{Choi et~al.}{2012}]{choi2012radiative}
Choi E.,  Ostriker J.~P.,  Naab T.,   Johansson P.~H.,  2012, ApJ, 754, 125

\bibitem[\protect\citeauthoryear{Choi, Brennan, Somerville, Ostriker, Hirschmann  \& Naab}{Choi et~al.}{2020}]{choi2020impact}
Choi E.,  Brennan R.,  Somerville R.~S.,  Ostriker J.~P.,  Hirschmann M.,   Naab T.,  2020, ApJ, 904, 8

\bibitem[\protect\citeauthoryear{Cid~Fernandes, Stasi{\'n}ska, Schlickmann, Mateus, Asari, Schoenell, Sodr{\'e}~Jr  \& collaboration)}{Cid~Fernandes et~al.}{2010}]{fernandes2010alternative}
Cid~Fernandes R.~C.,  Stasi{\'n}ska G.,  Schlickmann M.,  Mateus A.,  Asari N.~V.,  Schoenell W.,  Sodr{\'e}~Jr L.,   collaboration) S.,  2010, MNRAS, 403, 1036

\bibitem[\protect\citeauthoryear{Comerford, Pooley, Gerke  \& Madejski}{Comerford et~al.}{2011}]{comerford2011chandra}
Comerford J.~M.,  Pooley D.,  Gerke B.~F.,   Madejski G.~M.,  2011, ApJ, 737, L19

\bibitem[\protect\citeauthoryear{Comerford, Pooley, Barrows, Greene, Zakamska, Madejski  \& Cooper}{Comerford et~al.}{2015}]{comerford2015merger}
Comerford J.~M.,  Pooley D.,  Barrows R.~S.,  Greene J.~E.,  Zakamska N.~L.,  Madejski G.~M.,   Cooper M.~C.,  2015, ApJ, 806, 219

\bibitem[\protect\citeauthoryear{Comerford, Nevin, Stemo, M{\"u}ller-S{\'a}nchez, Barrows, Cooper  \& Newman}{Comerford et~al.}{2018}]{comerford2018origin}
Comerford J.~M.,  Nevin R.,  Stemo A.,  M{\"u}ller-S{\'a}nchez F.,  Barrows R.~S.,  Cooper M.~C.,   Newman J.~A.,  2018, AJ, 867, 66

\bibitem[\protect\citeauthoryear{Concas, Popesso, Brusa, Mainieri, Erfanianfar  \& Morselli}{Concas et~al.}{2017}]{concas2017light}
Concas A.,  Popesso P.,  Brusa M.,  Mainieri V.,  Erfanianfar G.,   Morselli L.,  2017, A\&A, 606, A36

\bibitem[\protect\citeauthoryear{Conroy, Wechsler  \& Kravtsov}{Conroy et~al.}{2006}]{conroy2006modeling}
Conroy C.,  Wechsler R.~H.,   Kravtsov A.~V.,  2006, ApJ, 647, 201

\bibitem[\protect\citeauthoryear{Costa, Rosdahl, Sijacki  \& Haehnelt}{Costa et~al.}{2018}]{costa2018driving}
Costa T.,  Rosdahl J.,  Sijacki D.,   Haehnelt M.~G.,  2018, MNRAS, 473, 4197

\bibitem[\protect\citeauthoryear{Croton et~al.,}{Croton et~al.}{2006}]{croton2006many}
Croton D.~J.,  et~al., 2006, MNRAS, 365, 11

\bibitem[\protect\citeauthoryear{DESI~Collaboration Aghamousa et~al.,}{DESI~Collaboration et~al.}{2016}]{aghamousa2016desi}
DESI~Collaboration Aghamousa A.,  et~al., 2016, arXiv preprint arXiv:1611.00036

\bibitem[\protect\citeauthoryear{DESI~Collaboration Abareshi et~al.,}{DESI~Collaboration et~al.}{2022}]{abareshi2022overview}
DESI~Collaboration Abareshi B.,  et~al., 2022, AJ, 164, 207

\bibitem[\protect\citeauthoryear{{DESI Collaboration} Abdul~Karim et~al.,}{{DESI Collaboration} et~al.}{2026}]{desi2026data}
{DESI Collaboration} Abdul~Karim M.,  et~al., 2026, AJ, 171, 285

\bibitem[\protect\citeauthoryear{Davis et~al.,}{Davis et~al.}{2022}]{davis2022radio}
Davis F.,  et~al., 2022, MNRAS, 511, 4109

\bibitem[\protect\citeauthoryear{Dekel \& Silk}{Dekel \& Silk}{1986}]{dekel1986origin}
Dekel A.,  Silk J.,  1986, ApJ, 303, 39

\bibitem[\protect\citeauthoryear{Del~Pino et~al.,}{Del~Pino et~al.}{2026}]{del2026ga}
Del~Pino B.~R.,  et~al., 2026, arXiv preprint arXiv:2601.06255

\bibitem[\protect\citeauthoryear{Dey et~al.,}{Dey et~al.}{2019}]{dey2019overview}
Dey A.,  et~al., 2019, AJ, 157, 168

\bibitem[\protect\citeauthoryear{Driver et~al.,}{Driver et~al.}{2009}]{driver2009gama}
Driver S.~P.,  et~al., 2009, Astronomy \& Geophysics, 50, 5

\bibitem[\protect\citeauthoryear{Eberhard, Reines, Gim, Darling  \& Greene}{Eberhard et~al.}{2025}]{eberhard2025dwarf}
Eberhard J.-M.,  Reines A.~E.,  Gim H.~B.,  Darling J.,   Greene J.~E.,  2025, ApJ, 978, 158

\bibitem[\protect\citeauthoryear{Efstathiou}{Efstathiou}{1992}]{efstathiou1992suppressing}
Efstathiou G.,  1992, MNRAS, 256, 43P

\bibitem[\protect\citeauthoryear{{Er{\'o}stegui}, {Mezcua}, {Siudek}, {Dom{\'\i}nguez S{\'a}nchez}  \& {Rodr{\'\i}guez Morales}}{{Er{\'o}stegui} et~al.}{2025}]{erostegui2025}
{Er{\'o}stegui} A.,  {Mezcua} M.,  {Siudek} M.,  {Dom{\'\i}nguez S{\'a}nchez} H.,   {Rodr{\'\i}guez Morales} V.,  2025, \mn@doi [\aap] {10.1051/0004-6361/202554387}, \href {https://ui.adsabs.harvard.edu/abs/2025A&A...699A.330E} {699, A330}

\bibitem[\protect\citeauthoryear{Ferland, Chatzikos, Guzm{\'a}n  et~al.}{Ferland et~al.}{2017}]{ferland2017rev}
Ferland G.,  Chatzikos M.,  Guzm{\'a}n F.,   et~al., 2017, Rev. Mex. Astron. Astrofis., 53, 385

\bibitem[\protect\citeauthoryear{Feruglio et~al.,}{Feruglio et~al.}{2015}]{feruglio2015multi}
Feruglio C.,  et~al., 2015, A\&A, 583, A99

\bibitem[\protect\citeauthoryear{Fierlinger, Burkert, Ntormousi, Fierlinger, Schartmann, Ballone, Krause  \& Diehl}{Fierlinger et~al.}{2016}]{fierlinger2016stellar}
Fierlinger K.~M.,  Burkert A.,  Ntormousi E.,  Fierlinger P.,  Schartmann M.,  Ballone A.,  Krause M.~G.,   Diehl R.,  2016, MNRAS, 456, 710

\bibitem[\protect\citeauthoryear{Fiore et~al.,}{Fiore et~al.}{2017}]{fiore2017agn}
Fiore F.,  et~al., 2017, A\&A, 601, A143

\bibitem[\protect\citeauthoryear{Flores, Mezcua  \& Morales}{Flores et~al.}{2025}]{flores2025manga}
Flores I.,  Mezcua M.,   Morales V.~R.,  2025, A\&A, 704, A267

\bibitem[\protect\citeauthoryear{Fritz, Franceschini  \& Hatziminaoglou}{Fritz et~al.}{2006}]{fritz2006revisiting}
Fritz J.,  Franceschini A.,   Hatziminaoglou E.,  2006, MNRAS, 366, 767

\bibitem[\protect\citeauthoryear{Girelli, Pozzetti, Bolzonella, Giocoli, Marulli  \& Baldi}{Girelli et~al.}{2020}]{girelli2020stellar}
Girelli G.,  Pozzetti L.,  Bolzonella M.,  Giocoli C.,  Marulli F.,   Baldi M.,  2020, A\&A, 634, A135

\bibitem[\protect\citeauthoryear{Greene, Strader  \& Ho}{Greene et~al.}{2020}]{greene2020intermediate}
Greene J.~E.,  Strader J.,   Ho L.~C.,  2020, ARA\&A, 58, 257

\bibitem[\protect\citeauthoryear{Guy et~al.,}{Guy et~al.}{2023}]{guy2023spectroscopic}
Guy J.,  et~al., 2023, AJ, 165, 144

\bibitem[\protect\citeauthoryear{Hahn et~al.,}{Hahn et~al.}{2023}]{hahn2023desi}
Hahn C.,  et~al., 2023, AJ, 165, 253

\bibitem[\protect\citeauthoryear{Harrison}{Harrison}{2017}]{harrison2017impact}
Harrison C.,  2017, Nat. Astronomy, 1, 0165

\bibitem[\protect\citeauthoryear{Harrison \& Ramos~Almeida}{Harrison \& Ramos~Almeida}{2024}]{harrison2024observational}
Harrison C.~M.,  Ramos~Almeida C.,  2024, Galaxies, 12, 17

\bibitem[\protect\citeauthoryear{Harrison et~al.,}{Harrison et~al.}{2012}]{harrison2012energetic}
Harrison C.,  et~al., 2012, MNRAS, 426, 1073

\bibitem[\protect\citeauthoryear{Harrison, Alexander, Mullaney  \& Swinbank}{Harrison et~al.}{2014}]{harrison2014kiloparsec}
Harrison C.,  Alexander D.,  Mullaney J.,   Swinbank A.,  2014, MNRAS, 441, 3306

\bibitem[\protect\citeauthoryear{Harrison, Costa, Tadhunter, Fl{\"u}tsch, Kakkad, Perna  \& Vietri}{Harrison et~al.}{2018}]{harrison2018agn}
Harrison C.,  Costa T.,  Tadhunter C.,  Fl{\"u}tsch A.,  Kakkad D.,  Perna M.,   Vietri G.,  2018, Nat. Astronomy, 2, 198

\bibitem[\protect\citeauthoryear{Hopkins, Kere{\v{s}}, O{\~n}orbe, Faucher-Gigu{\`e}re, Quataert, Murray  \& Bullock}{Hopkins et~al.}{2014}]{hopkins2014galaxies}
Hopkins P.~F.,  Kere{\v{s}} D.,  O{\~n}orbe J.,  Faucher-Gigu{\`e}re C.-A.,  Quataert E.,  Murray N.,   Bullock J.~S.,  2014, MNRAS, 445, 581

\bibitem[\protect\citeauthoryear{Hviding, Hainline, Rieke, Juneau, Lyu  \& Pucha}{Hviding et~al.}{2022}]{hviding2022new}
Hviding R.~E.,  Hainline K.~N.,  Rieke M.,  Juneau S.,  Lyu J.,   Pucha R.,  2022, AJ, 163, 224

\bibitem[\protect\citeauthoryear{Ivey et~al.,}{Ivey et~al.}{2026}]{ivey2026exploring}
Ivey L.,  et~al., 2026, MNRAS, 546, stag094

\bibitem[\protect\citeauthoryear{Jackson et~al.,}{Jackson et~al.}{2025}]{jackson2025diversity}
Jackson R.,  et~al., 2025, MNRAS, 539, 3797

\bibitem[\protect\citeauthoryear{Jarrett et~al.,}{Jarrett et~al.}{2011}]{jarrett2011spitzer}
Jarrett T.,  et~al., 2011, ApJ, 735, 112

\bibitem[\protect\citeauthoryear{Kauffmann et~al.,}{Kauffmann et~al.}{2003}]{kauffmann2003host}
Kauffmann G.,  et~al., 2003, MNRAS, 346, 1055

\bibitem[\protect\citeauthoryear{Kaviraj, Martin  \& Silk}{Kaviraj et~al.}{2019}]{kaviraj2019agn}
Kaviraj S.,  Martin G.,   Silk J.,  2019, MNRAS: Letters, 489, L12

\bibitem[\protect\citeauthoryear{Kennicutt~Jr \& Evans}{Kennicutt~Jr \& Evans}{2012}]{kennicutt2012star}
Kennicutt~Jr R.~C.,  Evans N.~J.,  2012, ARA\&A, 50, 531

\bibitem[\protect\citeauthoryear{Kewley, Dopita, Sutherland, Heisler  \& Trevena}{Kewley et~al.}{2001}]{kewley2001theoretical}
Kewley L.~J.,  Dopita M.,  Sutherland R.,  Heisler C.,   Trevena J.,  2001, ApJ, 556, 121

\bibitem[\protect\citeauthoryear{Kewley, Groves, Kauffmann  \& Heckman}{Kewley et~al.}{2006}]{kewley2006host}
Kewley L.~J.,  Groves B.,  Kauffmann G.,   Heckman T.,  2006, MNRAS, 372, 961

\bibitem[\protect\citeauthoryear{Kimm, Cen, Devriendt, Dubois  \& Slyz}{Kimm et~al.}{2015}]{kimm2015towards}
Kimm T.,  Cen R.,  Devriendt J.,  Dubois Y.,   Slyz A.,  2015, MNRAS, 451, 2900

\bibitem[\protect\citeauthoryear{Kormendy \& Ho}{Kormendy \& Ho}{2013}]{kormendy2013coevolution}
Kormendy J.,  Ho L.~C.,  2013, ARA\&A, 51, 511

\bibitem[\protect\citeauthoryear{Koudmani, Sijacki, Bourne  \& Smith}{Koudmani et~al.}{2019}]{koudmani2019fast}
Koudmani S.,  Sijacki D.,  Bourne M.~A.,   Smith M.~C.,  2019, MNRAS, 484, 2047

\bibitem[\protect\citeauthoryear{Koudmani, Henden  \& Sijacki}{Koudmani et~al.}{2021}]{koudmani2021little}
Koudmani S.,  Henden N.~A.,   Sijacki D.,  2021, MNRAS, 503, 3568

\bibitem[\protect\citeauthoryear{Koudmani, Sijacki  \& Smith}{Koudmani et~al.}{2022}]{koudmani2022two}
Koudmani S.,  Sijacki D.,   Smith M.~C.,  2022, MNRAS, 516, 2112

\bibitem[\protect\citeauthoryear{Koudmani, Rennehan, Somerville, Hayward, Angl{\'e}s-Alc{\'a}zar, Orr, Sands  \& Wellons}{Koudmani et~al.}{2025}]{koudmani2025diverse}
Koudmani S.,  Rennehan D.,  Somerville R.~S.,  Hayward C.~C.,  Angl{\'e}s-Alc{\'a}zar D.,  Orr M.~E.,  Sands I.~S.,   Wellons S.,  2025, MNRAS, 540, 1928

\bibitem[\protect\citeauthoryear{Lamastra, Bianchi, Matt, Perola, Barcons  \& Carrera}{Lamastra et~al.}{2009}]{lamastra2009bolometric}
Lamastra A.,  Bianchi S.,  Matt G.,  Perola G.~C.,  Barcons X.,   Carrera F.,  2009, A\&A, 504, 73

\bibitem[\protect\citeauthoryear{Law et~al.,}{Law et~al.}{2021}]{law2021sdss}
Law D.~R.,  et~al., 2021, ApJ, 915, 35

\bibitem[\protect\citeauthoryear{Leitherer et~al.,}{Leitherer et~al.}{1999}]{leitherer1999starburst99}
Leitherer C.,  et~al., 1999, ApJS, 123, 3

\bibitem[\protect\citeauthoryear{Lemons, Reines, Plotkin, Gallo  \& Greene}{Lemons et~al.}{2015}]{lemons2015x}
Lemons S.~M.,  Reines A.~E.,  Plotkin R.~M.,  Gallo E.,   Greene J.~E.,  2015, ApJ, 805, 12

\bibitem[\protect\citeauthoryear{Liu, Zakamska, Greene, Nesvadba  \& Liu}{Liu et~al.}{2013}]{liu2013observations}
Liu G.,  Zakamska N.~L.,  Greene J.~E.,  Nesvadba N.~P.,   Liu X.,  2013, MNRAS, 436, 2576

\bibitem[\protect\citeauthoryear{Liu, Veilleux, Canalizo, Rupke, Manzano-King, Bohn  \& Vivian}{Liu et~al.}{2020}]{liu2020integral}
Liu W.,  Veilleux S.,  Canalizo G.,  Rupke D.~S.,  Manzano-King C.~M.,  Bohn T.,   Vivian U.,  2020, ApJ, 905, 166

\bibitem[\protect\citeauthoryear{Lupi, Sbarrato  \& Carniani}{Lupi et~al.}{2020}]{lupi2020difficulties}
Lupi A.,  Sbarrato T.,   Carniani S.,  2020, MNRAS, 492, 2528

\bibitem[\protect\citeauthoryear{Luridiana, Morisset  \& Shaw}{Luridiana et~al.}{2015}]{luridiana2015pyneb}
Luridiana V.,  Morisset C.,   Shaw R.~A.,  2015, A\&A, 573, A42

\bibitem[\protect\citeauthoryear{Manzano-King \& Canalizo}{Manzano-King \& Canalizo}{2020}]{manzano2020active}
Manzano-King C.~M.,  Canalizo G.,  2020, MNRAS, 498, 4562

\bibitem[\protect\citeauthoryear{Manzano-King, Canalizo  \& Sales}{Manzano-King et~al.}{2019}]{manzano2019agn}
Manzano-King C.~M.,  Canalizo G.,   Sales L.~V.,  2019, ApJ, 884, 54

\bibitem[\protect\citeauthoryear{Mart{\'\i}n, Cob{\'a}, Cazzoli, Montero  \& Lavers}{Mart{\'\i}n et~al.}{2024}]{martin2024agn}
Mart{\'\i}n M.~V.,  Cob{\'a} C.~L.,  Cazzoli S.,  Montero E.~P.,   Lavers A.~C.,  2024, A\&A, 690, A397

\bibitem[\protect\citeauthoryear{Maschmann, Halle, Melchior, Combes  \& Chilingarian}{Maschmann et~al.}{2023}]{maschmann2023origin}
Maschmann D.,  Halle A.,  Melchior A.-L.,  Combes F.,   Chilingarian I.~V.,  2023, A\&A, 670, A46

\bibitem[\protect\citeauthoryear{Mateos et~al.,}{Mateos et~al.}{2012}]{mateos2012using}
Mateos S.,  et~al., 2012, MNRAS, 426, 3271

\bibitem[\protect\citeauthoryear{Mezcua}{Mezcua}{2017}]{mezcua2017observational}
Mezcua M.,  2017, International Journal of Modern Physics D, 26, 1730021

\bibitem[\protect\citeauthoryear{Mezcua \& Sanchez}{Mezcua \& Sanchez}{2024}]{mezcua2024manga}
Mezcua M.,  Sanchez H.~D.,  2024, MNRAS, 528, 5252

\bibitem[\protect\citeauthoryear{Mezcua, Civano, Fabbiano, Miyaji  \& Marchesi}{Mezcua et~al.}{2016}]{mezcua2016population}
Mezcua M.,  Civano F.,  Fabbiano G.,  Miyaji T.,   Marchesi S.,  2016, ApJ, 817, 20

\bibitem[\protect\citeauthoryear{Mezcua, Civano, Marchesi, Suh, Fabbiano  \& Volonteri}{Mezcua et~al.}{2018}]{mezcua2018intermediate}
Mezcua M.,  Civano F.,  Marchesi S.,  Suh H.,  Fabbiano G.,   Volonteri M.,  2018, MNRAS, 478, 2576

\bibitem[\protect\citeauthoryear{Mezcua, Suh  \& Civano}{Mezcua et~al.}{2019}]{mezcua2019radio}
Mezcua M.,  Suh H.,   Civano F.,  2019, MNRAS, 488, 685

\bibitem[\protect\citeauthoryear{Mezcua, Siudek, Suh, Valiante, Spinoso  \& Bonoli}{Mezcua et~al.}{2023}]{mezcua2023overmassive}
Mezcua M.,  Siudek M.,  Suh H.,  Valiante R.,  Spinoso D.,   Bonoli S.,  2023, ApJ, 943, L5

\bibitem[\protect\citeauthoryear{Miller et~al.,}{Miller et~al.}{2024}]{miller2024optical}
Miller T.~N.,  et~al., 2024, AJ, 168, 95

\bibitem[\protect\citeauthoryear{Moran, Shahinyan, Sugarman, V{\'e}lez  \& Eracleous}{Moran et~al.}{2014}]{moran2014black}
Moran E.~C.,  Shahinyan K.,  Sugarman H.~R.,  V{\'e}lez D.~O.,   Eracleous M.,  2014, AJ, 148, 136

\bibitem[\protect\citeauthoryear{Mori, Ferrara  \& Madau}{Mori et~al.}{2002}]{mori2002early}
Mori M.,  Ferrara A.,   Madau P.,  2002, ApJ, 571, 40

\bibitem[\protect\citeauthoryear{Moster, Somerville, Maulbetsch, Van Den~Bosch, Macci{\`o}, Naab  \& Oser}{Moster et~al.}{2010}]{moster2010constraints}
Moster B.~P.,  Somerville R.~S.,  Maulbetsch C.,  Van Den~Bosch F.~C.,  Macci{\`o} A.~V.,  Naab T.,   Oser L.,  2010, ApJ, 710, 903

\bibitem[\protect\citeauthoryear{Moster, Naab  \& White}{Moster et~al.}{2013}]{moster2013galactic}
Moster B.~P.,  Naab T.,   White S.~D.,  2013, MNRAS, 428, 3121

\bibitem[\protect\citeauthoryear{M{\"u}ller-S{\'a}nchez, Comerford, Nevin, Barrows, Cooper  \& Greene}{M{\"u}ller-S{\'a}nchez et~al.}{2015}]{muller2015origin}
M{\"u}ller-S{\'a}nchez F.,  Comerford J.~M.,  Nevin R.,  Barrows R.~S.,  Cooper M.~C.,   Greene J.~E.,  2015, ApJ, 813, 103

\bibitem[\protect\citeauthoryear{Navarro, Frenk  \& White}{Navarro et~al.}{1997}]{navarro1997universal}
Navarro J.~F.,  Frenk C.~S.,   White S.~D.,  1997, ApJ, 490, 493

\bibitem[\protect\citeauthoryear{Nevin, Comerford, M{\"u}ller-S{\'a}nchez, Barrows  \& Cooper}{Nevin et~al.}{2016}]{nevin2016origin}
Nevin R.,  Comerford J.,  M{\"u}ller-S{\'a}nchez F.,  Barrows R.,   Cooper M.,  2016, ApJ, 832, 67

\bibitem[\protect\citeauthoryear{Okamoto, Gao  \& Theuns}{Okamoto et~al.}{2008}]{okamoto2008mass}
Okamoto T.,  Gao L.,   Theuns T.,  2008, MNRAS, 390, 920

\bibitem[\protect\citeauthoryear{Osterbrock \& Ferland}{Osterbrock \& Ferland}{2006}]{osterbrock2006astrophysics}
Osterbrock D.~E.,  Ferland G.~J.,  2006, Astrophysics of gaseous nebulae and active galactic nuclei, 2nd

\bibitem[\protect\citeauthoryear{Penny et~al.,}{Penny et~al.}{2018}]{penny2018sdss}
Penny S.~J.,  et~al., 2018, MNRAS, 476, 979

\bibitem[\protect\citeauthoryear{Piotrowska, Bluck, Maiolino  \& Peng}{Piotrowska et~al.}{2022}]{piotrowska2022quenching}
Piotrowska J.~M.,  Bluck A.~F.,  Maiolino R.,   Peng Y.,  2022, MNRAS, 512, 1052

\bibitem[\protect\citeauthoryear{Polimera et~al.,}{Polimera et~al.}{2022}]{polimera2022resolve}
Polimera M.~S.,  et~al., 2022, ApJ, 931, 44

\bibitem[\protect\citeauthoryear{Poppett et~al.,}{Poppett et~al.}{2024}]{poppett2024overview}
Poppett C.,  et~al., 2024, AJ, 168, 245

\bibitem[\protect\citeauthoryear{Pucha et~al.,}{Pucha et~al.}{2025}]{pucha2025tripling}
Pucha R.,  et~al., 2025, ApJ, 982, 10

\bibitem[\protect\citeauthoryear{Pucha et~al.,}{Pucha et~al.}{2026}]{pucha2026new}
Pucha R.,  et~al., 2026, arXiv preprint arXiv:2606.02699

\bibitem[\protect\citeauthoryear{Qiu, McNamara, Bogdanovi{\'c}, Inayoshi  \& Ho}{Qiu et~al.}{2021}]{qiu2021mass}
Qiu Y.,  McNamara B.~R.,  Bogdanovi{\'c} T.,  Inayoshi K.,   Ho L.~C.,  2021, ApJ, 923, 256

\bibitem[\protect\citeauthoryear{Raichoor et~al.,}{Raichoor et~al.}{2023}]{raichoor2023target}
Raichoor A.,  et~al., 2023, AJ, 165, 126

\bibitem[\protect\citeauthoryear{Reines, Greene  \& Geha}{Reines et~al.}{2013}]{reines2013dwarf}
Reines A.~E.,  Greene J.~E.,   Geha M.,  2013, ApJ, 775, 116

\bibitem[\protect\citeauthoryear{Reines, Condon, Darling  \& Greene}{Reines et~al.}{2020}]{reines2020new}
Reines A.~E.,  Condon J.~J.,  Darling J.,   Greene J.~E.,  2020, ApJ, 888, 36

\bibitem[\protect\citeauthoryear{Rodr{\'\i}guez~Morales, Mezcua, Dom{\'\i}nguez~S{\'a}nchez, Audibert, M{\"u}ller-S{\'a}nchez, Siudek  \& Er{\'o}stegui}{Rodr{\'\i}guez~Morales et~al.}{2025}]{rodriguez2025manga}
Rodr{\'\i}guez~Morales V.,  Mezcua M.,  Dom{\'\i}nguez~S{\'a}nchez H.,  Audibert A.,  M{\"u}ller-S{\'a}nchez F.,  Siudek M.,   Er{\'o}stegui A.,  2025, A\&A, 697, A235

\bibitem[\protect\citeauthoryear{Rodr{\'\i}guez~del Pino, Arribas, Piqueras~L{\'o}pez, Villar-Mart{\'\i}n  \& Colina}{Rodr{\'\i}guez~del Pino et~al.}{2019}]{rodriguez2019properties}
Rodr{\'\i}guez~del Pino B.,  Arribas S.,  Piqueras~L{\'o}pez J.,  Villar-Mart{\'\i}n M.,   Colina L.,  2019, MNRAS, 486, 344

\bibitem[\protect\citeauthoryear{Rose, Tadhunter, Ramos~Almeida, Rodr{\'\i}guez~Zaur{\'\i}n, Santoro  \& Spence}{Rose et~al.}{2018}]{rose2018quantifying}
Rose M.,  Tadhunter C.,  Ramos~Almeida C.,  Rodr{\'\i}guez~Zaur{\'\i}n J.,  Santoro F.,   Spence R.,  2018, MNRAS, 474, 128

\bibitem[\protect\citeauthoryear{Rubin, Weiner, Koo, Martin, Prochaska, Coil  \& Newman}{Rubin et~al.}{2010}]{rubin2010persistence}
Rubin K.~H.,  Weiner B.~J.,  Koo D.~C.,  Martin C.~L.,  Prochaska J.~X.,  Coil A.~L.,   Newman J.~A.,  2010, ApJ, 719, 1503

\bibitem[\protect\citeauthoryear{Rupke, G{\"u}ltekin  \& Veilleux}{Rupke et~al.}{2017}]{rupke2017quasar}
Rupke D.~S.,  G{\"u}ltekin K.,   Veilleux S.,  2017, ApJ, 850, 40

\bibitem[\protect\citeauthoryear{Sacchi, Bogd{\'a}n, Chadayammuri  \& Ricarte}{Sacchi et~al.}{2024}]{sacchi2024x}
Sacchi A.,  Bogd{\'a}n {\'A}.,  Chadayammuri U.,   Ricarte A.,  2024, ApJ, 974, 14

\bibitem[\protect\citeauthoryear{Salehirad, Reines  \& Molina}{Salehirad et~al.}{2022}]{salehirad2022hundreds}
Salehirad S.,  Reines A.~E.,   Molina M.,  2022, ApJ, 937, 7

\bibitem[\protect\citeauthoryear{Salehirad, Reines  \& Molina}{Salehirad et~al.}{2025}]{salehirad2025ionized}
Salehirad S.,  Reines A.~E.,   Molina M.,  2025, ApJ, 979, 26

\bibitem[\protect\citeauthoryear{Sales, Wetzel  \& Fattahi}{Sales et~al.}{2022}]{sales2022baryonic}
Sales L.~V.,  Wetzel A.,   Fattahi A.,  2022, Nat. Astronomy, 6, 897

\bibitem[\protect\citeauthoryear{Sartori, Schawinski, Treister, Trakhtenbrot, Koss, Shirazi  \& Oh}{Sartori et~al.}{2015}]{sartori2015search}
Sartori L.~F.,  Schawinski K.,  Treister E.,  Trakhtenbrot B.,  Koss M.,  Shirazi M.,   Oh K.,  2015, MNRAS, 454, 3722

\bibitem[\protect\citeauthoryear{Schlafly et~al.,}{Schlafly et~al.}{2023}]{schlafly2023survey}
Schlafly E.~F.,  et~al., 2023, AJ, 166, 259

\bibitem[\protect\citeauthoryear{Schroetter et~al.,}{Schroetter et~al.}{2019}]{schroetter2019muse}
Schroetter I.,  et~al., 2019, MNRAS, 490, 4368

\bibitem[\protect\citeauthoryear{Scodeggio et~al.,}{Scodeggio et~al.}{2018}]{scodeggio2018vimos}
Scodeggio M.,  et~al., 2018, A\&A, 609, A84

\bibitem[\protect\citeauthoryear{Shankar, Lapi, Salucci, De~Zotti  \& Danese}{Shankar et~al.}{2006}]{shankar2006new}
Shankar F.,  Lapi A.,  Salucci P.,  De~Zotti G.,   Danese L.,  2006, ApJ, 643, 14

\bibitem[\protect\citeauthoryear{Sharma, Brooks, Somerville, Tremmel, Bellovary, Wright  \& Quinn}{Sharma et~al.}{2020}]{sharma2020black}
Sharma R.~S.,  Brooks A.~M.,  Somerville R.~S.,  Tremmel M.,  Bellovary J.,  Wright A.~C.,   Quinn T.~R.,  2020, ApJ, 897, 103

\bibitem[\protect\citeauthoryear{Sharma, Brooks, Tremmel, Bellovary, Ricarte  \& Quinn}{Sharma et~al.}{2022}]{sharma2022hidden}
Sharma R.~S.,  Brooks A.~M.,  Tremmel M.,  Bellovary J.,  Ricarte A.,   Quinn T.~R.,  2022, ApJ, 936, 82

\bibitem[\protect\citeauthoryear{Sharma, Brooks, Tremmel, Bellovary  \& Quinn}{Sharma et~al.}{2023}]{sharma2023active}
Sharma R.~S.,  Brooks A.~M.,  Tremmel M.,  Bellovary J.,   Quinn T.~R.,  2023, ApJ, 957, 16

\bibitem[\protect\citeauthoryear{Siudek, Mezcua  \& Krywult}{Siudek et~al.}{2023}]{siudek2023environment}
Siudek M.,  Mezcua M.,   Krywult J.,  2023, MNRAS, 518, 724

\bibitem[\protect\citeauthoryear{Siudek et~al.,}{Siudek et~al.}{2024}]{siudek2024value}
Siudek M.,  et~al., 2024, A\&A, 691, A308

\bibitem[\protect\citeauthoryear{Smethurst et~al.,}{Smethurst et~al.}{2021}]{smethurst2021kiloparsec}
Smethurst R.~J.,  et~al., 2021, MNRAS, 507, 3985

\bibitem[\protect\citeauthoryear{Smith, Sijacki  \& Shen}{Smith et~al.}{2019}]{smith2019cosmological}
Smith M.~C.,  Sijacki D.,   Shen S.,  2019, MNRAS, 485, 3317

\bibitem[\protect\citeauthoryear{Somerville, Hopkins, Cox, Robertson  \& Hernquist}{Somerville et~al.}{2008}]{somerville2008semi}
Somerville R.~S.,  Hopkins P.~F.,  Cox T.~J.,  Robertson B.~E.,   Hernquist L.,  2008, MNRAS, 391, 481

\bibitem[\protect\citeauthoryear{Stern et~al.,}{Stern et~al.}{2012}]{stern2012mid}
Stern D.,  et~al., 2012, ApJ, 753, 30

\bibitem[\protect\citeauthoryear{Su, Ji, Yuan, Xia  \& Zou}{Su et~al.}{2025}]{su2025positive}
Su T.,  Ji S.,  Yuan F.,  Xia H.,   Zou Y.,  2025, arXiv preprint arXiv:2510.20897

\bibitem[\protect\citeauthoryear{Vale \& Ostriker}{Vale \& Ostriker}{2006}]{vale2006non}
Vale A.,  Ostriker J.,  2006, MNRAS, 371, 1173

\bibitem[\protect\citeauthoryear{Vayner et~al.,}{Vayner et~al.}{2024}]{vayner2024first}
Vayner A.,  et~al., 2024, ApJ, 960, 126

\bibitem[\protect\citeauthoryear{Veilleux \& Osterbrock}{Veilleux \& Osterbrock}{1987}]{veilleux1987spectral}
Veilleux S.,  Osterbrock D.~E.,  1987, ApJS, 63, 295

\bibitem[\protect\citeauthoryear{Veilleux, Cecil  \& Bland-Hawthorn}{Veilleux et~al.}{2005}]{veilleux2005galactic}
Veilleux S.,  Cecil G.,   Bland-Hawthorn J.,  2005, ARA\&A, 43, 769

\bibitem[\protect\citeauthoryear{Wang, Li, Kauffmann  \& De~Lucia}{Wang et~al.}{2006}]{wang2006modelling}
Wang L.,  Li C.,  Kauffmann G.,   De~Lucia G.,  2006, MNRAS, 371, 537

\bibitem[\protect\citeauthoryear{Wang et~al.,}{Wang et~al.}{2025}]{wang2025rubies}
Wang B.,  et~al., 2025, AJ, 984, 121

\bibitem[\protect\citeauthoryear{Ward et~al.,}{Ward et~al.}{2022}]{ward2022variability}
Ward C.,  et~al., 2022, ApJ, 936, 104

\bibitem[\protect\citeauthoryear{Ward, Costa, Harrison  \& Mainieri}{Ward et~al.}{2024}]{ward2024agn}
Ward S.~R.,  Costa T.,  Harrison C.~M.,   Mainieri V.,  2024, Monthly Notices of the Royal Astronomical Society, 533, 1733

\bibitem[\protect\citeauthoryear{Wasleske, Baldassare  \& Carroll}{Wasleske et~al.}{2022}]{wasleske2022variable}
Wasleske E.~J.,  Baldassare V.~F.,   Carroll C.~M.,  2022, ApJ, 933, 37

\bibitem[\protect\citeauthoryear{Wellons et~al.,}{Wellons et~al.}{2023}]{wellons2023exploring}
Wellons S.,  et~al., 2023, MNRAS, 520, 5394

\bibitem[\protect\citeauthoryear{Wright et~al.,}{Wright et~al.}{2010}]{wright2010wide}
Wright E.~L.,  et~al., 2010, AJ, 140, 1868

\bibitem[\protect\citeauthoryear{Wylezalek, Flores, Zakamska, Greene  \& Riffel}{Wylezalek et~al.}{2020}]{wylezalek2020ionized}
Wylezalek D.,  Flores A.~M.,  Zakamska N.~L.,  Greene J.~E.,   Riffel R.~A.,  2020, MNRAS, 492, 4680

\bibitem[\protect\citeauthoryear{Xu et~al.,}{Xu et~al.}{2022}]{xu2022empress}
Xu Y.,  et~al., 2022, ApJ, 929, 134

\bibitem[\protect\citeauthoryear{Yang, Mo  \& van~den Bosch}{Yang et~al.}{2008}]{yang2008galaxy}
Yang X.,  Mo H.,   van~den Bosch F.~C.,  2008, ApJ, 676, 248

\bibitem[\protect\citeauthoryear{Yao et~al.,}{Yao et~al.}{2020}]{yao2020galaxy}
Yao H.,  et~al., 2020, ApJ, 903, 91

\bibitem[\protect\citeauthoryear{York et~al.,}{York et~al.}{2000}]{york2000sloan}
York D.~G.,  et~al., 2000, AJ, 120, 1579

\bibitem[\protect\citeauthoryear{Zheng, Shi, Bian, Yu, Wang, Chen, Li  \& Gu}{Zheng et~al.}{2023}]{zheng2023escaping}
Zheng Z.,  Shi Y.,  Bian F.,  Yu X.,  Wang J.,  Chen J.,  Li X.,   Gu Q.,  2023, MNRAS, 523, 3274

\bibitem[\protect\citeauthoryear{Zhuang \& Ho}{Zhuang \& Ho}{2023}]{zhuang2023evolutionary}
Zhuang M.-Y.,  Ho L.~C.,  2023, Nat. Astronomy, 7, 1376

\makeatother
\end{thebibliography}

 %para arXiv

\clearpage
\onecolumn
\begin{center}
{\Large \bfseries All Authors and Affiliations}
\end{center}
\vspace{0.3cm}

V. Rodr\'{i}guez Morales$^{1}$,
M. Mezcua$^{1,2}$,
Ragadeepika Pucha$^{3,4}$,
C. Circosta$^{5,6}$,
S. Juneau$^{7}$,
M. Siudek$^{1,8}$,
S. Panda$^{9}$,
J. Aguilar$^{10}$,
S. Ahlen$^{11}$,
D. Bianchi$^{12,13}$,
D. Brooks$^{5}$,
T. Claybaugh$^{10}$,
A. Cuceu$^{10}$,
A. de la Macorra$^{14}$,
B. Dey$^{15,16}$,
P. Doel$^{5}$,
J.E. Forero-Romero$^{17,18}$,
S. Gontcho A Gontcho$^{19}$,
G. Gutierrez$^{20}$,
J. Guy$^{10}$,
C. Hahn$^{21}$,
R. Joyce$^{7}$,
A. Kremin$^{10}$,
A. Lambert$^{10}$,
M. Landriau$^{10}$,
L. Le Guillou$^{22}$,
M. Manera$^{23,24}$,
P. Martini$^{25,26,27}$,
A. Meisner$^{7}$,
R. Miquel$^{24,28}$,
J. Moustakas$^{29}$,
W.J. Percival$^{30,31,32}$,
C. Poppett$^{7,33,34}$,
F. Prada$^{35}$,
I. P\'{e}rez-R\`{a}fols$^{36}$,
G. Rossi$^{37}$,
E. Sanchez$^{38}$,
D. Schlegel$^{10}$,
M. Schubnell$^{39,40}$,
D. Sprayberry$^{7}$,
G. Tarl\'{e}$^{40}$,
B.A. Weaver$^{7}$,
H. Zoo$^{41}$

\vspace{0.6cm}

\begin{enumerate}
\renewcommand{\labelenumi}{\arabic{enumi}.}
\item Institute of Space Sciences (ICE, CSIC), Campus UAB, Carrer de Magrans, 08193 Barcelona, Spain
\item Institut d'Estudis Espacials de Catalunya (IEEC), Edifici RDIT, Campus UPC, 08860 Castelldefels (Barcelona), Spain
\item Department of Physics and Astronomy, University of Utah, 115 South 1400 East, Salt Lake City, UT 84112, USA
\item Steward Observatory, University of Arizona, 933 North Cherry Avenue, Tucson, AZ 85719, USA
\item Department of Physics \& Astronomy, University College London, Gower Street, London, WC1E 6BT, UK
\item Institut de Radioastronomie Millim\'{e}trique (IRAM), 300 rue de la Piscine, 38400 Saint-Martin-d’Hères, France
\item NSF NOIRLab, 950 N. Cherry Avenue, Tucson, AZ 85719, USA
\item Instituto de Astrof\'{i}sica de Canarias, Calle V\'{i}a L\'{a}ctea, s/n, E-38205 La Laguna, Tenerife, Spain
\item International Gemini Observatory/NSF NOIRLab, Casilla 603, La Serena, Chile
\item Lawrence Berkeley National Laboratory, 1 Cyclotron Road, Berkeley, CA 94720, USA
\item Department of Physics, Boston University, 590 Commonwealth Avenue, Boston, MA 02215, USA
\item Dipartimento di Fisica ``Aldo Pontremoli'', Universit\`{a} degli Studi di Milano, Via Celoria 16, I-20133 Milano, Italy
\item INAF-Osservatorio Astronomico di Brera, Via Brera 28, 20122 Milano, Italy
\item Instituto de F\'{i}sica, Universidad Nacional Aut\'{o}noma de M\'{e}xico, Circuito de la Investigaci\'{o}n Cient\'{i}fica, Ciudad Universitaria, Cd. de M\'{e}xico, C.P. 04510, M\'{e}xico
\item Department of Astronomy \& Astrophysics, University of Toronto, Toronto, ON M5S 3H4, Canada
\item Department of Physics \& Astronomy and Pittsburgh Particle Physics, Astrophysics, and Cosmology Center (PITT PACC), University of Pittsburgh, 3941 O'Hara Street, Pittsburgh, PA 15260, USA
\item Departamento de F\'{i}sica, Universidad de los Andes, Cra. 1 No. 18A-10, Edificio Ip, CP 111711, Bogot\'{a}, Colombia
\item Observatorio Astron\'{o}mico, Universidad de los Andes, Cra. 1 No. 18A-10, Edificio H, CP 111711, Bogot\'{a}, Colombia
\item University of Virginia, Department of Astronomy, Charlottesville, VA 22904, USA
\item Fermi National Accelerator Laboratory, PO Box 500, Batavia, IL 60510, USA
\item Department of Astronomy, University of Texas at Austin, 2515 Speedway, TX 78712, USA
\item Sorbonne Universit\'{e}, CNRS/IN2P3, Laboratoire de Physique Nucl\'{e}aire et de Hautes Energies (LPNHE), FR-75005 Paris, France
\item Departament de F\'{i}sica, Serra H\'{u}nter, Universitat Aut\`{o}noma de Barcelona, 08193 Bellaterra (Barcelona), Spain
\item Institut de F\'{i}sica d’Altes Energies (IFAE), The Barcelona Institute of Science and Technology, Edifici Cn, Campus UAB, 08193, Bellaterra (Barcelona), Spain
\item Center for Cosmology and AstroParticle Physics, The Ohio State University, 191 West Woodruff Avenue, Columbus, OH 43210, USA
\item Department of Astronomy, The Ohio State University, 4055 McPherson Laboratory, 140 W 18th Avenue, Columbus, OH 43210, USA
\item The Ohio State University, Columbus, 43210 OH, USA
\item Instituci\'{o} Catalana de Recerca i Estudis Avan\c{c}ats, Passeig de Llu\'{i}s Companys, 23, 08010 Barcelona, Spain
\item Department of Physics and Astronomy, Siena University, 515 Loudon Road, Loudonville, NY 12211, USA
\item Department of Physics and Astronomy, University of Waterloo, 200 University Ave W, Waterloo, ON N2L 3G1, Canada
\item Perimeter Institute for Theoretical Physics, 31 Caroline St. North, Waterloo, ON N2L 2Y5, Canada
\item Waterloo Centre for Astrophysics, University of Waterloo, 200 University Ave W, Waterloo, ON N2L 3G1, Canada
\item Space Sciences Laboratory, University of California, Berkeley, 7 Gauss Way, Berkeley, CA 94720, USA
\item University of California, Berkeley, 110 Sproul Hall 5800, Berkeley, CA 94720, USA
\item Instituto de Astrof\'{i}sica de Andaluc\'{i}a (CSIC), Glorieta de la Astronom\'{i}a, s/n, E-18008 Granada, Spain
\item Departament de F\'{i}sica, EEBE, Universitat Polit\`{e}cnica de Catalunya, c/Eduard Maristany 10, 08930 Barcelona, Spain
\item Department of Physics and Astronomy, Sejong University, 209 Neungdong-ro, Gwangjin-gu, Seoul 05006, Republic of Korea
\item CIEMAT, Avenida Complutense 40, E-28040 Madrid, Spain
\item Department of Physics, University of Michigan, 450 Church Street, Ann Arbor, MI 48109, USA
\item University of Michigan, 500 S. State Street, Ann Arbor, MI 48109, USA
\item National Astronomical Observatories, Chinese Academy of Sciences, A20 Datun Road, Chaoyang District, Beijing, 100101, P.~R.~China

\end{enumerate}

\end{document}